\documentclass[nofootinbib,aps,pra,twocolumn,showkeys,groupaddress,preprintnumbers,floatfix]{revtex4-1}

\usepackage{dynlearn}
\usepackage{braket}
\newcommand{\ketbra}[1]{\ket{#1}\bra{#1}}

\newcommand{\rmE}{\mathrm{E}}
\newcommand{\rmD}{\mathrm{Df}}

\begin{document}

\def\ourTitle{
Thermodynamically-Efficient Local Computation and the\\
Inefficiency of Quantum Memory Compression}

\def\ourAbstract{
Modularity dissipation identifies how locally-implemented computation entails
costs beyond those required by Landauer's bound on thermodynamic computing. We
establish a general theorem for efficient local computation, giving the
necessary and sufficient conditions for a local operation to have zero
modularity cost. Applied to thermodynamically-generating stochastic processes
it confirms a conjecture that classical generators are efficient if and only if
they satisfy retrodiction, which places minimal memory requirements on the
generator. This extends immediately to quantum computation: Any quantum
simulator that employs quantum memory compression \emph{cannot} be
thermodynamically efficient.
}

\def\ourKeywords{
stochastic process, \texorpdfstring{\eM}{epsilon-machine}, predictor,
generator, causal states, modularity, Landauer principle.
}

\hypersetup{
  pdfauthor={James P. Crutchfield},
  pdftitle={\ourTitle},
  pdfsubject={\ourAbstract},
  pdfkeywords={\ourKeywords},
  pdfproducer={},
  pdfcreator={}
}

\author{Samuel P. Loomis}
\email{sloomis@ucdavis.edu}

\author{James P. Crutchfield}
\email{chaos@ucdavis.edu}
\affiliation{Complexity Sciences Center and Physics Department,
University of California at Davis, One Shields Avenue, Davis, CA 95616}

\date{\today}
\bibliographystyle{unsrt}

\title{\ourTitle}

\begin{abstract}
\ourAbstract
\end{abstract}

\keywords{\ourKeywords\\
DOI: \url{XX.XXXX/....}}

\preprint{\arxiv{2001.02258}}

\title{\ourTitle}
\date{\today}
\maketitle

\setstretch{1.1}

% \listoffixmes

\newcommand{\kB}{k_\text{B}}

\section{Introduction}

Recently, Google AI announced a breakthrough in quantum supremacy, using a
54-qubit processor (``Sycamore'') to complete a target computation in 200
seconds, claiming the world's fastest supercomputer would take more than 10,000
years to perform a similar computation \cite{Arut19}. Shortly afterward, IBM
announced that they had proven the Sycamore circuit could be successfully simulated on the
Summit supercomputer, leveraging its 250 PB storage and 200 petaFLOPS speed to
complete the target computation in a matter of days \cite{Pedn19}. This episode
highlights two important aspects of quantum computing: first, the importance of
memory and, second, the subtle relationship between computation and simulation. 

Feynman \cite{Feyn82a} broached the notion that quantum computers would be
singularly useful for the simulation of quantum processes, without supposing
that this would also make them advantageous at simulating classical processes.
Here, we explore issues raised by the recent developments in quantum computing,
focusing on the problem of simulating classical stochastic processes via
stochastic and quantum computers. We show that using quantum computers to
simulate classical processes typically requires nonzero thermodynamic cost,
while stochastic computers can theoretically achieve zero cost in simulating
classical processes. This supports the viewpoint originally put forth by
Feynman---that certain types of computers would each be advantageous at
simulating certain physical processes---which challenges the current claims of
quantum supremacy. Furthermore, we show that in both classical and quantum
simulations, thermodynamic efficiency places a lower bound on the required
memory of the simulator.

To demonstrate both, we must prove a new theorem
on the thermodynamic efficiency of local operations.
Correlation is a resource: it has been investigated
as such, in the formalism of {\em resource theories} \cite{Coec16a} such as 
that of local operations with classical communication \cite{Horo09a},
with public communication \cite{Maur93a}, and many others, as well as
the theory of local operations alone, under the
umbrella term of {\em common information} \cite{Gacs73a,Wyne75a,Kuma14a}.
Correlations have long been recognized as a thermal resource 
\cite{Land61a,Lloy89a,Boyd17c,Boyd17b},
enabling efficient computation to be performed
when taken properly into account.
Local operations that act only on
part of a larger system are known to never increase
the correlation between the part and the whole;
most often, they are destructive to correlations
and therefore resource-expensive. 

Thermodynamic dissipation induced by a local operation---say on system $A$ of
a bipartite system $AB$ to make a new joint system $CB$---is classically proportional to the difference in mutual informations \cite{Boyd17a}:
\begin{align*}
\Delta S_{\mathrm{loc}} = \kB T \left( I(A:B)-I(C:B) \right)
  ~.
\end{align*}
This can be asymptotically achieved for quantum systems \cite{Loom19a}.  By the
data processing inequality \cite{Cove06a,Niel10a}, it is always nonnegative:
$\Delta S_{\mathrm{loc}} \geq 0$. Optimal thermodynamic efficiency is achieved
when $\Delta S_{\mathrm{loc}}=0$.

To identify the conditions, in both classical and quantum computation, when
this is so, we draw from prior results on saturated information-theoretic
inequalities \cite{Petz86a,Petz88a,Rusk02a,Hayd04a,Moso04a,Moso05a,Petz08a}.
Specifically, using a generalized notion of quantum sufficient statistic
\cite{Jenc06,Petz08a,Lucz13a,Leif14a}, we show that a local operation on part
of a system is efficient if and only if it unitarily preserves the minimal
sufficient statistic of the part for the whole. Our geometric interpretation of
this also draws on recent progress on fixed points of quantum channels
\cite{Baum12a,Carb16a,Guan18a,Albe19a}.

Paralleling previous results on $\Delta S_{\mathrm{loc}}$ \cite{Boyd17a}, our
particular interest in locality arises from applying it to thermal
transformations that generate and manipulate stochastic processes. This is the
study of \emph{information engines} \cite{Mand12a,Boyd15a,Boyd17c,Boyd17b,Garn17b,Garn19a}.
Rooted in computational mechanics \cite{Crut88a,Shal98a,Trav12a,Crut12a}, which
investigates the inherent computational properties of natural processes and the
resources they consume, information engines embed stochastic processes and
Markovian generators in the physical world, where Landauer's bound for the cost
of erasure holds sway \cite{Land61a}.

A key result for information engines is the \emph{information-processing Second
Law} (IPSL): the cost of transforming one stochastic process into another by
any computation is at least the difference in their Kolmogorov-Sinai entropy
rates \cite{Boyd15a}. However, actual physical generators and transducers of
processes, with their own internal memory dynamics, often exceed the cost
required by the IPSL \cite{Boyd17a}. This arises from the temporal locality of
a physical generator---only operating from timestep-to-timestep, rather than
acting on the entire process at once. The additional dissipation $\Delta S_{\mathrm{loc}}$
induced by this temporal locality gives the true thermodynamic cost of
operating an information engine. 

Previous work explored optimal conditions for a classical information engine to
generate a process. Working from the hidden Markov model (HMM) \cite{Uppe97a}
that determines an engine's memory dynamics, it was conjectured that the HMM
must be \emph{retrodictive} to be optimal. For this to hold, the current memory state must be
a sufficient statistic of the \emph{future} data for predicting the \emph{past}
data \cite{Boyd17a}.

Employing a general result on conditions for reversible local computation, the
following confirms this conjecture, in the form of an equivalent condition on
the HMM's structure. We then extend this, showing that it holds for quantum
generators of stochastic processes
\cite{Gu12a,Maho16a,Riec16a,Bind17a,Agha17a,Agha18a,Thom18a,Loom18a,Liu19a,Loom19a}.
Notably, quantum generators are known to provide potentially unbounded
advantage in memory storage when compared to classical generators of the same
process \cite{Maho16a,Riec16a,Agha17a,Agha18a,Thom18a,Loom18a}. Surprisingly,
the advantage is contingent: optimally-efficient generators---those with $\Delta
S_{\mathrm{loc}}=0$---must not benefit from any memory compression.
We show this to be true not only for previously published
quantum generators, but for a new family of quantum generators
as well, derived from time reversal \cite{Croo08a,Crut08a,Elli11a,Thom18a}.

While important on its own, this also provides a complementary view to our
previous result on quantum generators which showed that a quantum-compressed
generator is never less thermodynamically-efficient than the classical
generator it compresses \cite{Loom19a}. Combined with our current result, one
concludes that a quantum-compressed generator is efficient with respect to the
generator it compresses but, to the extent that it is compressed, it cannot be
optimally efficient.  In short, only classical retrodictive generators achieve
the lower bound dictated by the IPSL. Practically, this highlights a pressing
need to experimentally explore the thermodynamics of quantum computing.

\section{Thermodynamics of Quantum Information Reservoirs}

The physical setting of our work is in the realm of \emph{information
reservoirs}---systems all of whose states have the same energy level.
Landauer's Principle for quantum systems says that to change an information
reservoir $A$ from state $\rho_A$ to state $\rho_A'$ requires a work cost
satisfying the lower bound:
\begin{align*}
  W \geq \kB T\ln 2\left(\H{\rho_A}-\H{\rho_A'}\right)
  ~.
\end{align*}
where $\H{\rho_A}$ is the von Neumann entropy \cite{Niel10a}.
Note that the lower bound $W_{\min} := \kB T\ln
2\left(\H{\rho_A}-\H{\rho_A'}\right)$ is simply the change in free energy
for an information reservoir.
Further, due to an information reservoir's trivial Hamiltonian, all of the
work $W$ becomes heat $Q$. Then the total entropy production---of system and
environment---is:
\begin{align*}
\Delta S & := Q + \kB T \ln 2\Delta \H{A} \\
  & = W - W_{\min}
  ~.
\end{align*}
Thus, not only does Landauer's Principle provide the lower bound, but reveals
that any work exceeding $W_{\min}$ represents dissipation.

Reference \cite{Boyd17a} showed that Landauer's bound may indeed be attained in
the quasistatic limit for any channel acting on a \emph{classical} information
reservoir. This result generally does not extend to single-shot quantum
channels \cite{Dahl11a}. However, when we consider asymptotically-many parallel
applications of a quantum channel, we recover the tightness of Landauer's bound
\cite{Loom19a}.

\begin{figure}[t]
\centering
\includegraphics[width=0.8\columnwidth]{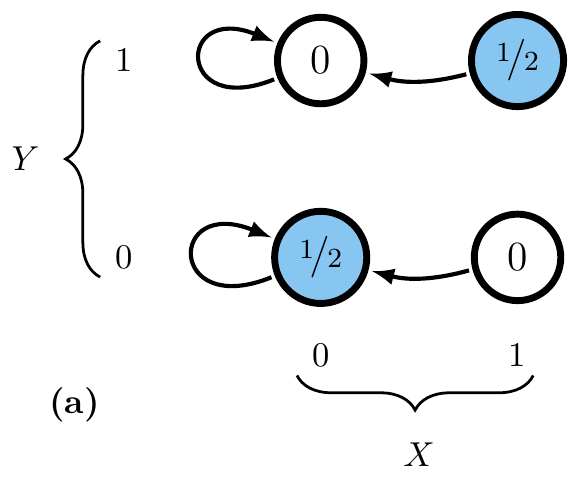}\\[25pt]
\includegraphics[width=0.8\columnwidth]{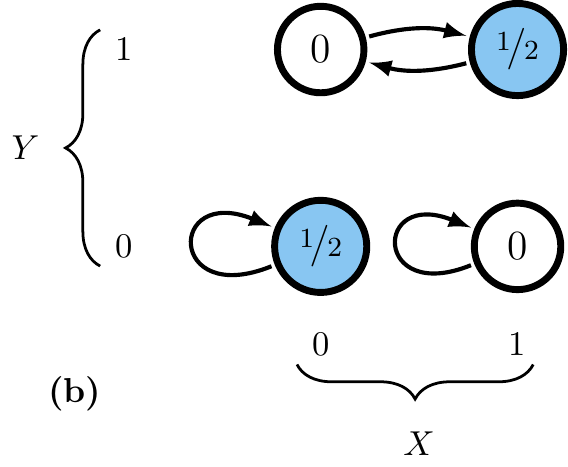}
\caption{Thermodynamics of locality: Suppose we have two bits
	$XY$ in a correlated state where $\half$ probability is in $XY=00$ and
	$\half$ probability is in $XY=11$. {\bf (a)} A thermodynamically
	irreversible operation can be performed to erase only $X$ (that is, set
	$X=0$ without changing $Y$) if we are not allowed to use knowledge about
	the state of $Y$.  {\bf (b)} A reversible operation can be performed to
	erase $X$ if we are allowed to use knowledge about $Y$. Both operations
	have the same outcome given our initial condition, but the nonlocal
	operation {\bf (a)} is more thermodynamically costly because it is
	irreversible. According to Thm. \ref{the:reversible_local}, operation {\bf
	(a)} is costly since it erases information in $X$ that is correlated with
	$Y$. 
    }
\label{fig:Locality}
\end{figure}

These statements are exceedingly general. To derive useful results, we must
place further constraints on the system dynamics to see how Landauer's bound is
affected. Reference \cite{Boyd17a} introduced the following perspective.
Consider a bipartite information reservoir $AB$, on which we wish to apply the
local channel $\mathcal{E}\otimes I_B$, where
$\mathcal{E}:\mathcal{B}\left(\mathcal{H}_A\right)\rightarrow
\mathcal{B}\left(\mathcal{H}_C\right)$ maps the states of system $A$ into those
of system $C$, transforming the initial joint state $\rho_{AB}$ to the final
state $\rho_{CB}$. The Landauer bound for $AB\rightarrow CB$ is given by
$W_{\min} = \kB T\ln 2\left(\H{\rho_{AB}}-\H{\rho_{CB}}\right)$. However,
since we constrained ourselves to use local manipulations, the lowest
achievable bound is actually $W_{\mathrm{loc}} := \kB T\ln
2\left(\H{\rho_{A}}-\H{\rho_{C}}\right)$. Thus, we must have dissipation of at
least:
\begin{align*}
\Delta S &\geq W_{\mathrm{loc}}-W_{\min} \\
  & = \kB T\ln 2
  \left(\H{\rho_{A}}-\H{\rho_{AB}}-\H{\rho_{C}}+\H{\rho_{CB}}\right) \\
  & = \kB T\ln 2
  \left(\mathrm{I}[A:B]-\mathrm{I}[C:B]\right)
  ~.
\end{align*}
where $\mathrm{I}[A:B] = \H{\rho_{A}}+\H{\rho_B}-\H{\rho_{AB}}$ is the quantum
mutual information. And so, we have a minimal \emph{locality dissipation}:
\begin{align*}
\Delta S_{\mathrm{loc}} := \kB T\ln 2
  \left(\mathrm{I}[A:B]-\mathrm{I}[C:B]\right)
  ~,
\end{align*}
which arises because we did not use the correlations to facilitate our erasure.
See Fig. \ref{fig:Locality} for a simple example of this phenomenon.

This local form of Landauer's Principle is still highly general, but the
following shows how to examine it for specific classical and quantum
computational architectures The key question we ask is: For which architectures
can $\Delta S_{\mathrm{loc}}$ be made to exactly vanish?  We first we consider
this problem generally and then provide a solution.

\section{Reversible Local Computation}

Suppose we are given a bipartite system $AB$ with state $\rho_{AB}$.  We wish
to determine the conditions for a local channel $\mathcal{E}_A\otimes I_B$ that
maps $A$ to $C$:
\begin{align*}
\rho_{CB}' = \mathcal{E}_A\otimes I_B\left(\rho_{AB}\right)
\end{align*}
to preserve the mutual information $I(A:B) = I(C:B)$.
Proofs of the following results are provided in the Supplementary Material.

Stating our result requires first defining the quantum notion of a sufficient
statistic. Previously, quantum sufficient statistics of $A$ for $B$ were
defined when $AB$ is a classical-quantum state \cite{Leif14a}. That is, when
$\rho_{AB}$ commutes with a local measurement on $A$. They were also introduced
in the setting of sufficient statistics for a family of states
\cite{Jenc06,Petz08a}. This corresponds to the case where $AB$ is
quantum-classical---$\rho_{AB}$ commutes with a local measurement on $B$. Our
definition generalizes these cases to fully-quantal correlations between $A$
and $B$.

We start, as an example, by giving the following definition of a minimum
sufficient statistic of a classical joint random variable $XY\sim \Pr(x,y)$ in
terms of an equivalence relation. We define the predictive equivalence relation
$\sim$ for which $x\sim x'$ if and only if $\Pr(y|x)=\Pr(y|x')$ for all $y$.
The {\em minimum sufficient statistic} (MSS) $[X]_Y$ is given by the
equivalence classes $[x]_Y:=\{x':x\sim x'\}$. Let us denote $\Sigma := [X]_Y$
and let $\Pr(y|\sigma):=\Pr(y|x)$ for any $x\in\sigma$.

This cannot be {\em directly} generalized to the quantum setting since
correlations between $A$ and $B$ cannot always be described in the form of
states conditioned on the outcome of a local measurement on $A$. If the
latter were the case, the state would be classical-quantum, but general quantum
correlations can be much more complicated than these. However, we can take the
most informative local measurement that does not disturb $\rho_{AB}$ and then
consider the ``atomic'' quantum correlations it leaves behind.

Let $\rho_{AB}$ be a bipartite quantum state. A \emph{maximal local commuting
measurement} (MLCM) of $A$ for $B$ is any local measurement $X$ with projectors $\{\Pi^{(x)}\}$
on system $A$
such that:
\begin{align*}
  \rho_{AB} = \bigoplus_{x} \Pr(X=x) \rho^{(x)}_{AB}
  ~,
\end{align*}
where:
\begin{align*}
\Pr(X=x) = \mathrm{Tr}\left((\Pi^{(x)}_X\otimes I_B) \rho_{AB}\right)
\end{align*}
and:
\begin{align*}
\Pr(X=x)\rho^{(x)}_{AB}
  = (\Pi^{(x)}_X\otimes I_B) \rho_{AB}(\Pi^{(x)}_X\otimes I_B)
  ~,
\end{align*}
and any further local measurement $Y$ on $\rho^{(x)}_{AB}$ disturbs the state:
\begin{align*}
\rho^{(x)}_{AB}
  \neq \sum_y (\Pi^{(y)}_Y\otimes I_B)\rho^{(x)}_{AB}(\Pi^{(y)}_Y\otimes I_B)
  ~.
\end{align*}
We call the states $\left\{ \rho^{(x)}_{AB}\right\}$ \emph{quantum correlation atoms}.

\begin{Prop}[MLCM uniqueness]
Given a state $\rho_{AB}$, there is a unique MLCM of $A$ for $B$.
\end{Prop} 

Now, as in the classical setting, we define an equivalence class over the
values of the MLCM via the equivalence between their quantum correlation atoms.
Classically, these atoms are simply the conditional probability distributions
$\Pr(\cdot|x)$; in the classical-quantum setting, they are the conditional
quantum states $\rho^{(x)}_B$. Note that each is defined as a distribution on
the variable $Y$ or system $B$. In contrast, the general quantum correlation
atoms $\rho^{(x)}_{AB}$ depend on both systems $A$ and $B$.

The resulting challenge is resolved in the following way. Let $\rho_{AB}$ be a
bipartite quantum state and let $X$ be the MLCM of $A$ for $B$. We define the
{\em correlation equivalence} relation $x\sim x'$ over values of $X$ where
$x\sim x'$ if and only if $\rho^{(x)}_{AB} = (U\otimes I_B)\rho^{(x')}_{AB}
(U^{\dagger}\otimes I_B)$ for a local unitary $U$. 
  
Finally, we define the \emph{Minimal Local Sufficient Statistic} (MLSS)
$[X]_{\sim}$ as the equivalence class $[x]_{\sim}:=\{x':x'\sim x\}$ generated
by the relation $\sim$ between correlation atoms. Thus, our notion of
sufficiency of $A$ for $B$ is to find the most informative local measurement
and then coarse-grain its outcomes by unitary equivalence over their
correlation atoms. The correlation atoms and the MLSS $[X]_{\sim}$ together
describe the correlation structure of the system $AB$.

The machinery is now in place to state our result.
The proof depends on previous results regarding the fixed points
of stochastic channels \cite{Baum12a,Carb16a,Guan18a,Albe19a}
and saturated information-theoretic inequalities \cite{Petz86a,Petz88a,Rusk02a,Hayd04a,Moso04a,Moso05a,Petz08a}.
This background and the proof are described in the SM.

\begin{The}[Reversible local operations]
\label{the:reversible_local}
Let $\rho_{AB}$ be a bipartite quantum state and let $\mathcal{E}_A\otimes I_B$
be a local operation with
$\mathcal{E}_A:\mathcal{B}(\mathcal{H}_A)\rightarrow\mathcal{B}(\mathcal{H}_C)$.
Suppose $X$ is the MLCM of $\rho_{AB}$ and $Y$, that of $\rho_{CB} =
\mathcal{E}_A\otimes I_B\left(\rho_{AB}\right)$. The decomposition into
correlation atoms is:
\begin{align}
\label{eq:decomp}
\rho_{AB} & = \bigoplus_{x} \Pr{}_{A}\left(x\right) \rho_{A B}^{(x)}
  ~\text{and}\\
\rho_{CB} & = \bigoplus_{y} \Pr{}_{C}\left(y\right) \rho_{C B}^{(y)}
  ~.
\end{align}
Then, $I\left(A:B\right) = I\left(C:B\right)$ if and only if $\mathcal{E}_A$ can
be expressed by Kraus operators of the form:
\begin{align}
\label{eq:kraus_decomp}
K^{(\alpha)}
  = \bigoplus_{x,y} e^{i \phi_{xy\alpha}}\sqrt{\Pr(y,\alpha|x)} U^{(y|x)}
  ~,
\end{align}
where $\phi_{xy\alpha}$ is any arbitrary phase and
$\Pr\left(y,\alpha\middle|x\right)$ is a stochastic channel that is nonzero
only when $\rho_{A B}^{(x)}$ and $\rho_{C B}^{(y)}$ are equivalent up to a
local unitary operation $U^{(y|x)}$ that maps $\mathcal{H}^{(x)}_A$ to
$\mathcal{H}^{(y)}_C$.
\end{The}

The theorem's classical form follows as a corollary.

\begin{Cor}[Reversible local operations, classical]
\label{the:reversible_local_class}
Let $XY$ be a joint random variable and let $\Pr(Z=z|X=x)$ be a channel from
$\mathcal{X}$ to some set $\mathcal{Z}$, resulting in the joint random variable
$ZY$. Then $I(X:Y)=I(Z:Y)$ if and only if $\Pr(Z=z|X=x)>0$ only when
$\Pr(Y=y|Z=z)=\Pr(Y=y|X=x)$ for all $y$.
\end{Cor}

In light of the previous section,
there is a simple thermodynamic interpretation of Thm.
\ref{the:reversible_local} and Cor. \ref{the:reversible_local_class}: local
channels that circumvent dissipation due to their locality 
(\ie, those which have $\Delta S_{\mathrm{loc}}=0$) 
are precisely those
channels that preserve the sufficiency structure of the joint state. They may
create and destroy any information that is not stored in the sufficient
statistic and the correlation atoms. However, the sufficient statistic itself
must be conserved and the correlation atoms must be only unitarily transformed.

We now turn to apply this perspective to classical and quantum
\emph{generators}---systems that use thermodynamic mechanisms to produce
stochastic processes. We compute the necessary and sufficient conditions for
these generators to have zero locality dissipation: $\Delta
S_{\mathrm{loc}}=0$.  And so, in this way we determine precise criteria for
when they are thermodynamically efficient.

\section{Thermodynamics of Classical Generators}
\label{sm:class-app}

A classical generator is the physical implementation
of a hidden Markov model (HMM) \cite{Uppe97a}
$\mathfrak{G}=(\mathcal{S},\mathcal{X},\{T_{s's}^{(x)}\})$, where (here)
$\mathcal{S}$ is countable, $\mathcal{X}$ is finite, and for each
$x\in\mathcal{X}$, $\mathbf{T}^{(x)}$ is a matrix with values given by a
stochastic channel from $\mathcal{S}$ to $\mathcal{S}\times \mathcal{Y}$,
${T}^{(x)}_{s's} := \Pr{}_{\mathfrak{G}}(s',x|s)$. We define generators to use
\emph{recurrent} HMMs, which means the total transition matrix $T_{s's}:=\sum_x
T^{(x)}_{s's}$ is irreducible. In this case, there is a unique stationary
distribution $\pi_{\mathfrak{G}}(s)$ over states $\mathcal{S}$ satisfying
$\pi_{\mathfrak{G}}(s)>0$, $\sum_s \pi_{\mathfrak{G}}(s)=1$, and
$\sum_{s}T_{s's}\pi_{\mathfrak{G}}(s) = \pi_{\mathfrak{G}}(s')$.

\begin{figure}[t]
\centering
\includegraphics[width=\columnwidth]{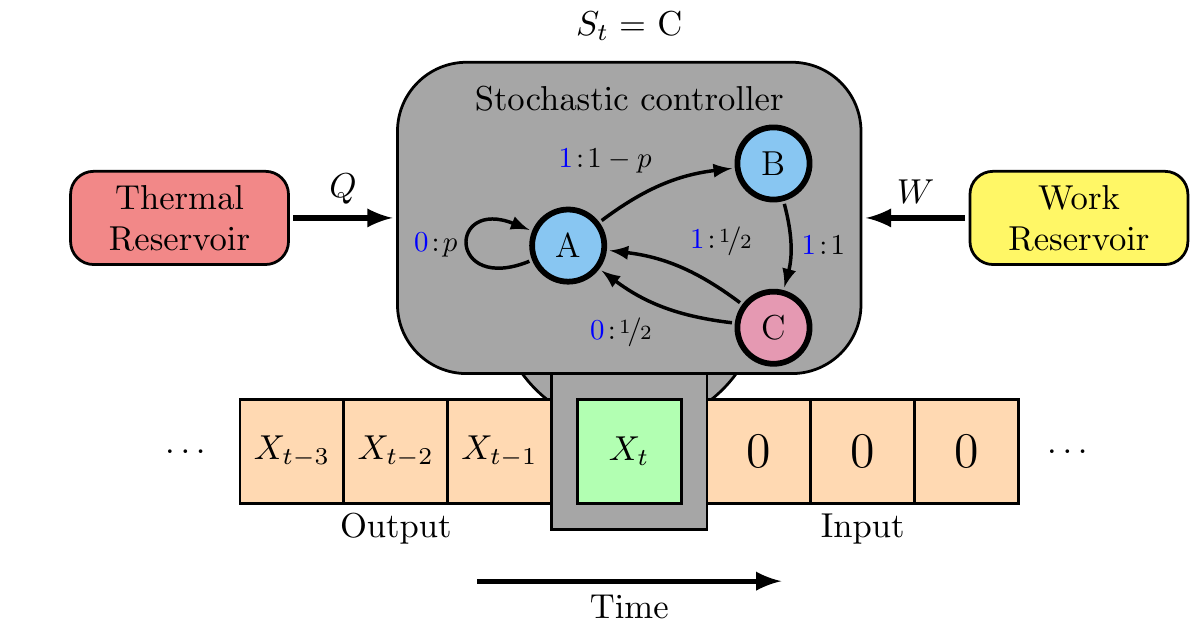}
\caption{{\bf Information ratchet sequentially generating a symbol string on an
	empty tape}: At time step $t$, $S_t$ is the random variable for the ratchet
	state.  The generated symbols in the generated (output) process are denoted
	by $X_{t-1},X_{t-2},X_{t-3}, \ldots$. The most recently generated symbol
	$X_t$ (green) is determined by the internal dynamics of the ratchet's
	memory, using heat $Q$ from the thermal reservoir as well as work $W$ from
	the work reservoir. \emph{(Inside ratchet)}: Ratchet memory dynamics and
	symbol emission are governed by the conditional probabilities
	$\Pr(s_{t+1},x_t|s_t)$, where $s_t$ is the current state at time $t$, $x_t$
	is the generated symbol, and $s_{t+1}$ is the new state. Graphically, this
	is represented by a hidden Markov model, depicted here as a
	state-transition diagram in which nodes are states $s$ and edges represent
	transitions $s\rightarrow s'$ labeled by the generated symbol and
	associated probability: ${\color{blue} x}:\Pr(s',x|s)$. (Reproduced with
	permission from Ref. \protect\cite{Loom19a}.)
    }
  \label{fig:Ratchet}
\end{figure}

During its operation, a generator's function is to produce a \emph{stochastic
process}---for each $\ell$, a probability distribution $\Pr{}_{\mathfrak{G}}\left(x_1\dots
x_\ell\right)$ over words $x_1\dots x_\ell\in \mathcal{X}^\ell$. The probabilities for
words of length $\ell$ generated by $\mathfrak{G}$ are defined by:
\begin{align*}
\Pr{}_{\mathfrak{G}}\left(x_1\dots x_\ell\right)
  := \sum_{s_0\dots s_\ell\in\mathcal{S}^{\ell+1}} 
  T^{(x_\ell)}_{s_\ell s_{\ell-1}}\dots T^{(x_1)}_{s_1 s_0} \pi(s_0)
  ~.
\end{align*}
Typically, we view a generator as operating over discrete time, writing out a
sequence of symbols from $x\in\mathcal{X}$ on a tape, while internally
transforming its memory state; see \cref{fig:Ratchet}. Starting with an initial
state $S_0\sim \pi(s)$ and empty tape at time $t=0$, the entire system at time
$t$ is described by the joint random variable $X_1\dots X_t S_t$, with
distribution:
\begin{align*}
\Pr{}_{\mathfrak{G}}\left(x_1\dots x_{t}, s_t\right)
  := \sum_{s_0\dots s_{t-1}\in\mathcal{S}^{t}} 
  T^{(x_t)}_{s_t s_{t-1}}\dots T^{(x_1)}_{s_1 s_0} \pi_{\mathfrak{G}}(s_0)
  ~.
\end{align*}
Continuing this technique, one can compute the joint random variable $X_1\dots
X_t S_t X_{t+1} S_{t+1}$.

This picture of a generator as operating on a tape while continually erasing and
rewriting its internal memory allows us to define the possible thermodynamics,
also shown in \cref{fig:Ratchet}. Erasure generally requires work, drawn from the
work reservoir, while the creation of noise often allows the extraction of work,
which is represented in our sign convention by drawing negative work from the
reservoir. Producing a process $X_1\dots X_t \sim \Pr\left(x_1\dots x_{t}\right)$
of length $t$ has an associated work cost $W \geq - \kB T \ln 2 H\left(X_1\dots
X_t\right)$. The negative sign, as discussed, indicates work $\kB T \ln 2
~H\left(X_1\dots X_t\right)$ may be transferred from the thermal reservoir to the
work reservoir. For large $t$, this can be asymptotically expressed by the work
rate $W/t \geq - \kB T \ln 2 ~\hmu$, where:
\begin{align*}
  \hmu := \lim_{t\rightarrow \infty} \frac{1}{t} \H{X_1\dots X_t}
\end{align*}
is the process' \emph{Kolmogorov-Sinai entropy rate} \cite{Boyd15a}. This is a
reasonable description of the average entropy rate of a process that is
\emph{stationary}---that is, $\Pr(X_t \dots X_{t+\ell}=x_1\dots x_{\ell-1})$ is
independent of $t$---and \emph{ergodic}. Said differently, for large $t$ a
typical realization $x_1\dots x_t$ contains the word $\widehat{x}_1\dots
\widehat{x}_\ell$ approximately $t \times \Pr(\widehat{x}_1\dots
\widehat{x}_\ell)$ times. Recurrent generators produce exactly these sorts of
processes.

Now, a given generator cannot necessarily be implemented as efficiently as the
minimal work rate $W_{\min} := - \kB T \ln 2 ~\hmu$ indicates. This is because a
generator acts temporally locally, only being able to use its current memory
state to generate the next memory state and symbol. The true cost at time $t$
must be bounded below by $W_{\mathrm{loc}} := W_{\min}+\Delta S_{\mathrm{loc}}$,
where in this case the locality dissipation is \cite{Boyd17a}:
\begin{align*}
\Delta S_{\mathrm{loc}} =& \kB T\ln 2
	~(I\left(S_t :X_{1}\dots X_{t} \right) \\
  & \qquad - I\left(S_{t+1} X_{t+1}:X_{1}\dots X_{t}\right))
  ~.
\end{align*}
In this case, the dissipation does not represent work lost to heat but rather the
increase in tape entropy that did not facilitate converting heat into work.
To understand this in some detail, this section identifies the necessary and
sufficient conditions for efficient generators---those with $\Delta
S_{\mathrm{loc}}=0$.

To state our result for classical generators, we must introduce two further
notions regarding generators. As before,
proofs of results are given in the SM.
Consider a partition of $\mathcal{S}$:
$\mathcal{P}=\left\{ \mathcal{P}_\theta\right\}$, $\mathcal{P}_\theta\cap
\mathcal{P}_{\theta'}$, $\bigcup_\theta \mathcal{P}_\theta = \mathcal{S}$,
labeled by index $\theta$. Let:
\begin{align*}
\Pr{}_{\mathfrak{G}^{\mathcal{P}}}\left(\theta',x|\theta\right)
  := \sum_{\substack{s'\in\mathcal{P}_{\theta'}\\ s\in\mathcal{P}_\theta}} 
  \Pr{}_{\mathfrak{G}}\left(s',x|s\right) \pi(s|\theta)
  ~,
\end{align*}
with $\pi_{\mathfrak{G}}(s|\theta) =
\pi_{\mathfrak{G}}(s)/\pi_{\mathfrak{G}^{\mathcal{P}}}(\theta)$ and
$\pi_{\mathfrak{G}^{\mathcal{P}}}(\theta) = \sum_{s\in\mathcal{P}_\theta}
\pi_{\mathfrak{G}}(s)$. We say a partition $\left\{ \mathcal{P}_\theta\right\}$
is \emph{mergeable} with respect to the generator
$\mathfrak{G}=(\mathcal{S},\mathcal{X},\{T_{s's}^{(x)}\})$ if the merged
generator
$\mathfrak{G}^{\mathcal{P}}=(\mathcal{P},\mathcal{X},\{\widetilde{T}_{\theta'\theta}^{(x)}\})$,
with transitions $\widetilde{T}_{\theta'\theta}^{(x)}:=
\Pr\left(\theta',x|\theta\right)$, generates the same process as the original.

Pertinent to our goals here is the notion of \emph{retrodictive equivalence}.
Let $\Pr{}_{\mathfrak{G}}\left(x_1\dots x_t|s_t\right) :=
\Pr{}_{\mathfrak{G}}\left(x_1\dots x_t,s_t\right) /\pi{}_{\mathfrak{G}}(s_t)$.
Given two states $s,s'\in\mathcal{S}$ of a generator
$(\mathcal{S},\mathcal{X},\{T_{s's}^{(x)}\})$, we say that $s\sim s'$ if
$\Pr{}_{\mathfrak{G}}\left(x_1\dots
x_t|s\right)=\Pr{}_{\mathfrak{G}}\left(x_1\dots x_t|s'\right)$ for all words
$x_1\dots x_t$. The equivalence class $[S_t]_\sim$ is the sufficient statistic of
$S_t$ for predicting the past symbols $X_1\dots X_t$.  The set $\mathcal{P}_\sim
:= \{[s]_\sim:s\in\mathcal{S}\}$ of equivalence classes is a partition on
$\mathcal{S}$ that we index by $\sigma$. 

\begin{Prop}
\label{the:mergeable}
Given a generator $(\mathcal{S},\mathcal{X},\{T_{s's}^{(x)}\})$, the partition
$\mathcal{P}:=\mathcal{P}_\sim$ induced by retrodictive equivalence is
mergeable.
\end{Prop}

We now state our theorem for efficient classical generators:

\begin{The}\label{the:class_retro}
A generator $\mathfrak{G}=(\mathcal{S},\mathcal{X},\{T_{s's}^{(x)}\})$
satisfies $I\left(S_t :X_{1}\dots X_{t} \right)= I\left(S_{t+1}
X_{t+1}:X_{1}\dots X_{t}\right)$ for all $t$ if and only if the retrodictively
state-merged generator
$\mathfrak{G}^{\mathcal{P}}=(\mathcal{P}_{\sim},\mathcal{X},\{\widetilde{T}_{\sigma'\sigma}^{(x)}\})$
satisfies $\widetilde{T}_{\sigma'\sigma}^{(x)}\propto \delta_{\sigma,f(\sigma',x)}$
for some function $f:\mathcal{S}\times\mathcal{X}\rightarrow \mathcal{S}$.
\end{The}

We say that a generator
$\mathfrak{G}=(\mathcal{S},\mathcal{X},\{T_{s's}^{(x)}\})$ satisfying
$T_{s's}^{(x)}\propto \delta_{s,f(s',x)}$ for some $f$ is \emph{co-unifilar}.
The dual property $T_{s's}^{(x)}\propto \delta_{s',f(s,x)}$ for some $f$ is
called \emph{unifilar} \cite{Ash65a}. For every process, there is a unique
generator, called the \emph{reverse \eM}, constructed by retrodictively
state-merging any co-unifilar generator \cite{Loom18a}. Similarly, using a
different partition, called \emph{predictive} equivalence on states, any
unifilar generator for a process can be state-merged into a unique generator
called the \emph{forward \eM} of that process \cite{Loom18a}.

The reverse \eM has the following property. Let $\overrightarrow{X}_t:= X_{t+1}
X_{t+1}\dots$ represent all future generated symbols, the reverse \eM state
$\Sigma_t$ at time $t$ is the minimum sufficient statistic of
$\overrightarrow{X}_t$ for predicting $X_1\dots X_t$. Any generator whose state
$S_t$ is a sufficient statistic of $\overrightarrow{X}_t$ for $X_1\dots X_t$ is
called a {\em retrodictor}. The reverse \eM can then be considered the minimal
retrodictor.

Reference \cite{Boyd17a} conjectured that the necessary and sufficient
condition for $\Delta S_{\mathrm{loc}}=0$ is that the generator in question is
a retrodictor. In the SM we confirm this by establishing that the conditions of
Thm. \ref{the:class_retro} imply the generator is a retrodictor.

A similar result, for classical generators, was presented in \cite{Garn17b}
where a lower bound on $\Delta S_{\mathrm{loc}}$ was derived for predictive
generators (Eq. (A23) in \cite{Garn17b}). A consequence of this bound is that
$\Delta S_{\mathrm{loc}}=0$ only when the predictor is also a retrodictor.
However, this bound does not extend to nonpredictive generators. In contrast,
Thm. \ref{the:class_retro} applies to all generators.

Our result is complemented by another recent result \cite{Garn19a}, which
demonstrated how from a predictive generator one can construct a sequence of
generators that asymptotically approach a retrodictor and whose dissipation
$\Delta S_{\mathrm{loc}}$ asymptotically approaches zero. Helpfully, this
result points to possible perturbative extensions of Thm. \ref{the:class_retro}.

These results bear on the trade-off between dissipation and \emph{memory} for
classical generators. The reverse (forward) \eM, being a state-merging of any
co-unifilar (unifilar) generator, is minimal with respect to the co-unifilar
(unifilar) generators via all quantifications of the memory, such as the number
of memory states $|\mathcal{S}|$ and the entropy $\H{S}$ \cite{Loom18a}.

As a consequence, we now see that the above showed that any
\emph{thermodynamically efficient} generator can be state-merged into a
co-unifilar generator. This means it can be further state-merged into the
reverse \eM of the process it generates. In short, thermodynamic efficiency
comes with a memory constraint. And, when the memory falls below this
constraint, \emph{dissipation must be present}.

\section{Thermodynamics of Quantum Machines}

A process' forward \eM, a key player in the previous section, may be concretely
defined as the unique generator $\mathfrak{G} =
(\mathcal{S},\mathcal{X},\{T_{s's}^{(x)}\})$ for a given process satisfying \cite{Trav12a}:
\begin{enumerate}
\setlength{\topsep}{-5pt}
\setlength{\itemsep}{-5pt}
\setlength{\parsep}{-5pt}
\item \emph{Recurrence}: $T_{s's}:=\sum_x T^{(x)}_{s's}$ is an irreducible matrix;
\item \emph{Unifilarity}: $T_{s's}^{(x)}>0$ only when  $s'=f(s,x)$ for some
	function $f:\mathcal{S}\times\mathcal{X}\rightarrow \mathcal{S}$;
\item \emph{Predictively Distinct States}:
  $\Pr(x_t x_{t+1}\dots x_{\ell}|s_t)=\Pr(x_t x_{t+1}\dots x_{\ell}|s_t')$
  for all $\ell$ and $x_t x_{t+1}\dots x_{\ell}$ implies $s_t=s_{t'}$.
\end{enumerate}
\EMs are a process' minimal unifilar generators, in the sense that they are
smallest with respect to the number of memory states $|\mathcal{S}|$, the
entropy $\H{S}$, and all other ways of measuring memory, such as the R{\'e}nyi
entropies $\mathrm{H}_\alpha[S]:=\frac{1}{1-\alpha}\log_2\left(\sum_s
\pi_{\mathfrak{G}}(s)^{\alpha}\right)$. In this, they are unique.

However, one can implement \eMs with even lower memory costs, by encoding them
in a quantum system and generating symbols by means of a noisy measurement.
This encoding is called a \emph{$q$-machine}. In terms of qubits, as a unit of
size, these implementations can generate the same process at a much lower
memory cost that the \eM's bit-based memory cost. It has also been shown that
these quantum implementations have a lower locality cost $W_{\mathrm{loc}}$
than their corresponding \eM, and so they are more thermodynamically efficient
\cite{Loom19a}.

This section identifies the constraints for quantum generators to have zero
dissipation; that is, $\Delta S_{\mathrm{loc}}=0$. We show that this results in
a peculiar pair of constraints. First, the forward \eM memory must not be
smaller than the memory of the reverse \eM. (This mirrors the results of Thm.
\ref{the:class_retro} in SM.) Second, the quantum generator achieves no
compression. That is, the memory of the quantum generator in qubits is
precisely the memory of the forward \eM in bits. Thus, compression of memory
and perfect thermodynamic efficiency are exclusive outcomes. 

To state this precisely, we review $q$-machines and introduce several new
definitions to capture their properties. (See the SM for the proofs.)

Given a forward \eM
$\mathfrak{G}=(\mathcal{S},\mathcal{X},\{T_{s's}^{(x)}\})$, for any set of phases
$\{\phi_{xs}:x\in\mathcal{X},s\in\mathcal{S}\}$ there is an encoding
$\{\ket{\psi_s}:s\in\mathcal{S}\}$ of the memory states $\mathcal{S}$ into a
Hilbert space $\mathcal{H}_S$ and a set of Kraus operators
$\{K^{(x)}:x\in\mathcal{X}\}$ on said Hilbert space such that:
\begin{align*}
K^{(x)}\ket{\psi_s}
  = e^{i\phi_{xs}}\sqrt{T^{(x)}_{f(s,x),s}}\ket{\psi_{f(s,x)}}
  ~.
\end{align*}
This expression implicitly defines the Kraus operators given the encoding
$\{\ket{\psi_s}\}$. The encoding, in turn, is determined up to a unitary
transformation by the following constraint on their overlaps:
\begin{align*}
\braket{\psi_r|\psi_s} = \sum_{x\in\mathcal{X}}
  e^{i(\phi_{xs}-\phi_{xr})}\sqrt{T^{(x)}_{r',r}T^{(x)}_{s',s}}
  \braket{\psi_{r'}|\psi_{s'}}
  ~,
\end{align*}
where $r'=f(r,x)$ and $s'=f(s,x)$. This equation has a unique solution for every
choice of phases $\{\phi_{xs}\}$ \cite{Liu19a}. 

We note that if $\pi_{\mathfrak{G}}(s)$ is the \eM's stationary distribution,
then the stationary state of this quantum generator is given by the ensemble:
\begin{align*}
\rho_\pi= \sum_{s} \pi_{\mathfrak{G}}(s)\ketbra{\psi_s}
\end{align*}
and satisfies:
\begin{align*}
\rho_\pi = \sum_x K^{(x)} \rho_\pi K^{(x)\dagger}
  ~.
\end{align*}
When we say that a quantum generator \emph{uses less memory than its classical
counterpart}, we mean that $\dim \mathcal{H}_S \leq |\mathcal{S}|$, $\H{\rho_\pi}
\leq \H{S}$, and further that $\mathrm{H}_\alpha[\rho_\pi] \leq
\mathrm{H}_\alpha[S]$, where
$\mathrm{H}_\alpha[\rho_\pi]:=\frac{1}{1-\alpha}\log_2\mathrm{Tr}\left[\rho_\pi^\alpha\right]$
are the R{\'e}nyi-von Neumann entropies \cite{Maho16a,Thom18a,Loom18a}.

To see this quantum generator as a physical system, as in \cref{fig:Ratchet},
requires us interpreting the tape being written on as a series of copies of a
single Hilbert space $\mathcal{H}_A$ that represents one cell on the tape.  On
$\mathcal{H}_A$ we define the computational basis $\{\ket{x}:x\in\mathcal{X}\}$
in which outputs will be written. The system at time $t$ can be described using
the joint Hilbert space $\mathcal{H}_{A_1}\otimes \mathcal{H}_{A_t}\otimes
\mathcal{H}_S$, where each $\mathcal{H}_{A_k}$ is unitarily equivalent to
$\mathcal{H}_A$, and the state is:
\begin{align*}
  \rho_{\mathfrak{G}}(t) := \sum_{x_1\dots x_t} \ketbra{x_1\dots x_t}\otimes K^{(x_t\dots x_1)}\rho_\pi K^{(x_t\dots x_1)\dagger}
  ~,
\end{align*}
where $K^{(x_t\dots x_1)} = K^{(x_t)}\dots K^{(x_1)}$ and $\ket{x_1\dots x_t} =
\ket{x_1}_{A_1}\otimes\dots\otimes \ket{x_t}_{A_t}$. From this we get the
process generated by the \eM and quantum generator in terms of the Kraus
operators as:
\begin{align}
\Pr{}_{\mathfrak{G}}\left(x_1\dots x_t\right)
  := \mathrm{Tr}\left[K^{(x_t\dots x_1)}\rho_\pi K^{(x_t\dots x_1)\dagger}\right]
  ~.
\label{eq:qgen}
\end{align}

Let us now briefly discuss the thermodynamic properties of quantum generators,
homing in on our main result about conditions for their efficiency. The
previous section discussed how a process, to be generated, requires the minimal
work rate $W_{\min} = - \kB T \ln 2 ~\hmu$. However, this is not typically
achievable for classical generators. The same principle holds for quantum
generators: Since they act temporally locally, the true cost at time $t$ is
bounded below by $W_{\mathrm{loc}} = W_{\min}+\Delta S_{\mathrm{loc}}$ and the
locality dissipation $\Delta S_{\mathrm{loc}}$ has the same form:
\begin{align}
  \begin{split}
  \Delta S_{\mathrm{loc}} &= \kB T\ln 2 ~(I\left(S_t :A_{1}\dots A_{t} \right) \\
  & \qquad - I\left(S_{t+1} A_{t+1}:A_{1}\dots A_{t}\right))
  ~. 
  \end{split}
\label{eq:quantum-diss}
\end{align}
There are two crucial differences, though. First, the mutual information $I$
above is the \emph{quantum} mutual information derived from the von Neumann
entropy. Second, even the work rate $W_{\mathrm{loc}}$ is not necessarily
achievable in the single-shot case \cite{Dahl11a}. However, it may be attained
for asymptotically parallel generation \cite{Loom19a}. We will not concern
ourselves with this second problem here. Our intent is to focus, as in the
previous section, on the necessary and sufficient conditions for $\Delta
S_{\mathrm{loc}}=0$. 

The preceding material was, in fact, review. We now introduce a simple
partition that may be constructed on the memory states of the \eM for a given
quantum implementation. Specifically, we define the \emph{maximal commuting
partition} (MCP) on $\mathcal{S}$ to be the most refined partition
$\{\mathcal{B}_\theta\}$ such that the overlap matrix $\braket{\psi_r|\psi_s}$
is block-diagonal. That is, $\{\mathcal{B}_\theta\}$ is such that
$\braket{\psi_r|\psi_s}=0$ if $r\in \mathcal{B}_\theta$ and
$s\in\mathcal{B}_{\theta'}$ for $\theta\neq \theta'$.

Our result on thermodynamically-efficient quantum generators is as follows.

\begin{The}[Maximally-efficient quantum generator]
\label{the:quant_pre}
Let $\mathfrak{G}=(\mathcal{S},\mathcal{X},\{T_{s's}^{(x)}\})$ be a given
process' \eM. Suppose we build from it a quantum generator with encoding
$\left\{\psi_s\right\}$ and Kraus operators $\{K^{(x)}\}$. Let
$\mathcal{B}:=\{\mathcal{B}_\theta\}$ be the MCP of $\mathcal{S}$. Then the
quantum generator has $\Delta S_{\mathrm{loc}}=0$ if and only if the partition
$\mathcal{B}$ is trivially maximal---in that $|\mathcal{B}_\theta|=1$ for each
$\theta$---and the retrodictively state-merged generator
$\mathfrak{G}^{\mathcal{B}}$ of $\mathfrak{G}$ is co-unifilar.
\end{The}

\begin{figure}[t]
\centering
\includegraphics[width=\columnwidth]{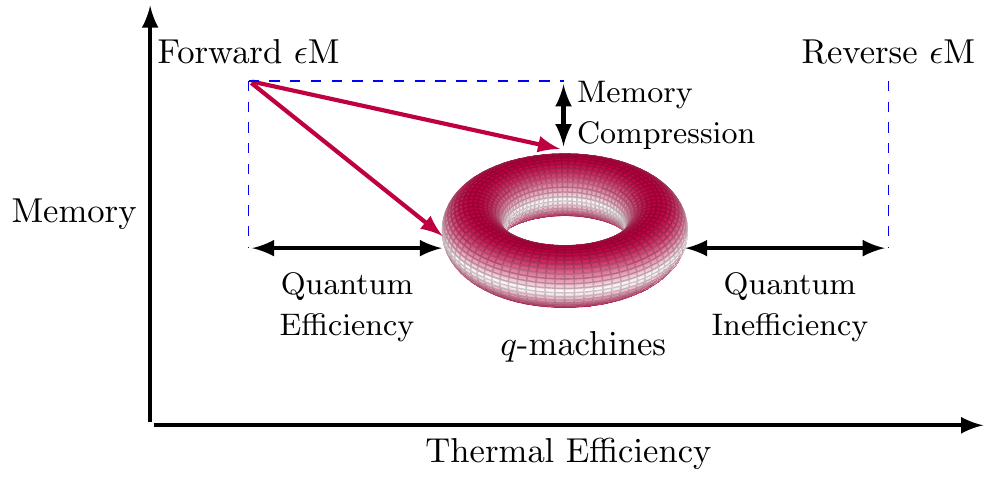}
\caption{Performance trade-offs for $q$-machines, whose variety and dependence
	on phases $\{\phi_{xs}\}$ is depicted by a torus: Under all ways of
	quantifying memory, the $q$-machines constructed from a predictor achieve
	nonnegative memory compression \cite{Loom18a}, and they also have a smaller
	dissipation $\Delta S_{\mathrm{loc}}$, rendering them more
	thermodynamically efficient \cite{Loom19a}.  However, to achieve positive
	compression, they must also have a nonzero $\Delta S_{\mathrm{loc}}$,
	rendering them less efficient than a classical retrodictor.
	}
\label{fig:ForwardQmach}
\end{figure}

We previously found that, in the limit of asymptotically parallel generation, a
quantum generator is always more thermodynamically efficient than its
corresponding \eM, in that it has a lower dissipation \cite{Loom19a}. Yet this
does not imply dissipation can be made to vanish for quantum generators of a
process. In fact, only for processes whose forward \eM is also a retrodictor
can dissipation be made to vanish. In these cases, the memory states will be
orthogonally encoded, and so no memory compression is achieved, which is seen
by the trivial maximality of $\mathcal{B}$. The situation is heuristically
represented in Fig. \ref{fig:ForwardQmach}.

\section{Thermodynamics of reverse $q$-machines}

We showed that forward \eMs compressed via the $q$-machine cannot achieve the
efficiency of a classical retrodictor. However, one may wonder what happens to
a retrodictor's optimal efficiency if it is directly compressed. We now
demonstrate a method for such compression, derived from the time-reversal of
the $q$-machine, and prove that even here any nonzero compression of memory
precludes optimal efficiency.

A process' reverse \eM may be defined similarly to the forward \eM as the
unique generator $\mathfrak{G} = (\mathcal{S},\mathcal{X},\{T_{s's}^{(x)}\})$
for a given process satisfying:
\begin{enumerate}
\setlength{\topsep}{-5pt}
\setlength{\itemsep}{-5pt}
\setlength{\parsep}{-5pt}
\item \emph{Recurrence}: $T_{s's}:=\sum_x T^{(x)}_{s's}$ is an irreducible matrix;
\item \emph{Co-unifilarity}: $T_{s's}^{(x)}>0$ only when  $s=f(s',x)$ for some
	function $f:\mathcal{S}\times\mathcal{X}\rightarrow \mathcal{S}$;
\item \emph{Retrodictively Distinct States}:
	$\Pr(x_1 \dots x_{t}|s_t)=\Pr(x_1 \dots x_{t}|s_t')$
	for all $t$ and $x_1 \dots x_{t}$ implies $s_t=s_{t'}$.
\end{enumerate}
Reverse \eMs are a process' minimal co-unifilar generators, in the sense that
they are smallest with respect to the number of memory states $|\mathcal{S}|$,
the entropy $\H{S}$, and all other ways of measuring memory, such as the
R{\'e}nyi entropies $\mathrm{H}_\alpha[S]:=\frac{1}{1-\alpha}\log_2\left(\sum_s
\pi_{\mathfrak{G}}(s)^{\alpha}\right)$.

There is an intricate relationship between forward and reverse \eMs that can
only be appreciated in the language of time reversal. The time-reverse of a
generator $\mathfrak{G} = (\mathcal{S},\mathcal{X},\{T_{s's}^{(x)}\})$ is the
generator
$\widetilde{\mathfrak{G}}=(\mathcal{S},\mathcal{X},\{\widetilde{T}{}_{s's}^{(x)}\})$
where $\widetilde{T}{}_{s's}^{(x)} = \pi_s T_{s's}^{(x)}/\pi_{s'}$ \cite{Elli11a}.
The generator $\widetilde{G}$ is associated with the reverse process,
$\Pr_{\widetilde{G}}\left(x_1 \dots x_t\right)=\Pr_{{G}}\left(x_t \dots
x_1\right)$. Note that time reversal preserves both the state space
$\mathcal{S}$ and the stationary distribution $\pi_s$.

Given a process' forward \eM $\mathfrak{F}$, its time reverse
$\widetilde{\mathfrak{F}}$ is the \emph{reverse} \eM of the \emph{reverse}
process.  Conversely, given a process' reverse \eM $\mathfrak{G}$, its time
reverse $\widetilde{\mathfrak{G}}$ is the \emph{forward} \eM of the
\emph{reverse} process. Since the stationary distribution and state space are
preserved under time reversal, $\mathfrak{F}$ and $\widetilde{\mathfrak{F}}$
have the same memory costs, as do $\mathfrak{G}$ and
$\widetilde{\mathfrak{G}}$. However, somewhat surprisingly, this does not mean
that $\mathfrak{F}$ and $\mathfrak{G}$ have the same memory costs
\cite{Crut08a}.

Previous work compared the results of compressing the forward \eM
$\mathfrak{F}$ of a process {\em and} the forward \eM
$\widetilde{\mathfrak{G}}$ of the reverse process using the $q$-machine
formalism. The result, for compressing $\widetilde{\mathfrak{G}}$, is a
$q$-machine that generates the reverse process---remarkably, with identical
cost to the $q$-machine constructed from $\mathfrak{F}$ \cite{Thom18a}.

The $q$-machine constructed from $\widetilde{\mathfrak{G}}$ is a quantum
process and as such can itself undergo quantum time-reversal \cite{Croo08a},
resulting in a new process that we call the \emph{reverse $q$-machine}. Just as
the $q$-machine compresses $\widetilde{\mathfrak{G}}$, the reverse $q$-machine
is a compression of $\mathfrak{G}$. Though the reverse $q$-machine is derived
from the $q$-machine via time-reversal, there is genuinely new physics present,
as the dissipation $\Delta S_{\mathrm{loc}}$ (\cref{eq:quantum-diss}) is not
invariant under time-reversal. Thus, they must be approached as a separate case
from the traditional $q$-machine when examining their thermodynamic efficiency.

The details of the time-reversal are handled in the SM. Here, we present the
resulting technique for compressing the reverse \eM. Given a reverse \eM
$\mathfrak{G} = (\mathcal{S},\mathcal{X},\{T_{s's}^{(x)}\})$, for any set of
phases $\{\phi_{xs}:x\in\mathcal{X},s\in\mathcal{S}\}$ there is an encoding
$\{\ket{\psi_s}:s\in\mathcal{S}\}$ of {\em orthogonal} states into a Hilbert
space $\mathcal{H}_S$ and a set of Kraus operators
$\{K^{(x)}:x\in\mathcal{X}\}$ on said Hilbert space such that:
\begin{align*}
  K^{(x)}\ket{\psi_s} = \sum_{s'\in\mathcal{S}}
  e^{i\phi_{xs'}}\sqrt{T^{(x)}_{s's}}\ket{\psi_{s'}}
  ~.
\end{align*}
The orthogonality of $\{\ket{\psi_s}\}$ allows us to turn this into an explicit
definition of the Kraus operators:
\begin{align*}
  K^{(x)} = \sum_{s'\in\mathcal{S}}
  e^{i\phi_{xs'}}\sqrt{T^{(x)}_{s'f(s',x)}}\ket{\psi_{s'}}\bra{\psi_{f(s',x)}}
  ~.
\end{align*}
The stationary state $\rho_\pi$ of this machine is, unlike the $q$-machine,
generically not expressible as an ensemble of the encoding states
$\{\ket{\psi_s}\}$. If this were so, the orthogonality of $\{\ket{\psi_s}\}$
would make them a diagonalizing basis for $\rho_\pi$, and we would achieve no
memory compression. Rather, compression is achieved for the reverse $q$-machine
precisely because the stationary state $\rho_\pi$ is generically not diagonal
in the encoding states---in contrast to the $q$-machine, which derived
compression from the nonorthogonality of the encoding states.

The reverse $q$-machine stochastic dynamics \cref{eq:qgen} and thermodynamics
\cref{eq:quantum-diss} are defined precisely as those for $q$-machines in the
previous section. As before, to prove our result we must define a special
partition of the generator states. Here, it is important to note a relationship
between a process' forward \eM
$\mathfrak{F}=\left(\mathcal{P},\mathcal{X},\left\{R^{(x)}_{p'p}\right\}\right)$
and its reverse \eM
$\mathfrak{G}=\left(\mathcal{S},\mathcal{X},\left\{T^{(x)}_{s's}\right\}\right)$.
Specifically, the state $S_t$ of $\mathfrak{G}$ after seeing the word $x_1\dots
x_t$ and the state $P_t$ of $\mathfrak{F}$ after the same are related by:
\begin{align*}
  \Pr{}_{\widetilde{\mathfrak{G}}}\left(s_t |{x}_1 \dots {x}_t \right)
  = \sum_{p_t} \Pr{}_{\mathcal{C}}\left(s_t|p_t\right)
  \Pr{}_{{\mathfrak{F}}}\left(p_t |{x}_1 \dots {x}_t \right)
\end{align*}
for some channel $\Pr{}_{\mathcal{C}}\left(s|p\right)$. Let $\lambda_p$ be the
stationary distribution of $\mathfrak{F}$'s states and let
$\Pr{}_{\mathcal{E}}\left(s'|s\right) = \sum_p
\Pr{}_{\mathcal{C}}\left(s|p\right)\Pr{}_{\mathcal{C}}\left(s'|p\right)\lambda_p
/ \pi_s$. Let $\mathcal{B}=\{\mathcal{B}_\theta\}$ be the ergodic partition of
$\Pr{}_{\mathcal{E}}\left(s'|s\right)$, such that
$\Pr{}_{\mathcal{E}}\left(s'|s\right)>0$ only when $\theta(s)=\theta(s')$. The
SM shows that the $\rho_\pi$ is diagonal in the blocks defined by $\mathcal{B}$.

Our result for reverse $q$-machines, proven in the SM, can now be stated:

\begin{The}[Maximally-efficient reverse $q$-machine]
\label{the:quant_retro}
Let $\mathfrak{G}=(\mathcal{S},\mathcal{X},\{T_{s's}^{(x)}\})$ be a given
process' reverse \eM. Suppose we build from it a reverse $q$-machine with
encoding $\left\{\ket{\psi_s}\right\}$ and Kraus operators $\{K^{(x)}\}$. Let
$\mathcal{B}:=\{\mathcal{B}_\theta\}$ be the MCP of $\mathcal{S}$. Then the
reverse $q$-machine has $\Delta S_{\mathrm{loc}}=0$ if and only if the
partition $\mathcal{B}$ is trivially maximal---in that $|\mathcal{B}_\theta|=1$
for each $\theta$---and the predictively state-merged generator
$\mathfrak{G}^{\mathcal{B}}$ of $\mathfrak{G}$ is unifilar.
\end{The}

\begin{figure}[t]
\centering
\includegraphics[width=\columnwidth]{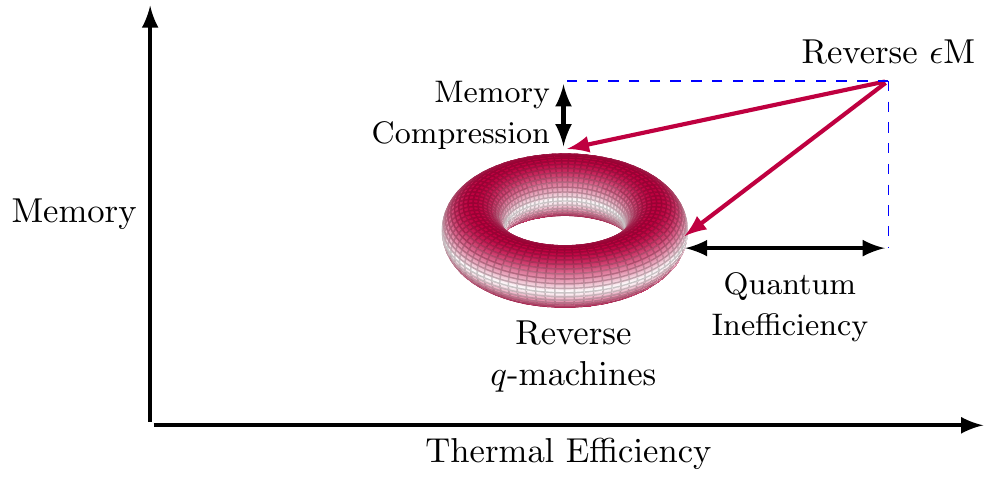}
\caption{Performance trade-offs for reverse $q$-machines, whose variety and
	dependence on $\{\phi_{xs}\}$ is represented by a torus: Under all
	quantifications of memory, the reverse $q$-machines constructed from a
	retrodictor achieve nonnegative memory compression. However, to achieve
	positive compression, they must also have a nonzero dissipation $\Delta
	S_{\mathrm{loc}}$. The latter renders them less thermodynamically
	efficient. 
	}
\label{fig:ReverseQmach}
\end{figure}

Notice that this is a similar statement to that made in the last section and is
essentially its time reverse. It implies that the only reverse \eMs which can
be quantally compressed are those which are also predictive generators. Also,
again the trivial maximality of the ergodic partition $\mathcal{B}$ implies an
inability to achieve nonzero compression. A heuristic diagram of the situation
is shown in Fig. \ref{fig:ReverseQmach}.

In conjunction with the previous section, this is a profound result on the
efficiency of quantum memory compression. Distinct from the classical case,
where Thm. \ref{the:class_retro} established that \emph{every} process has
certain generators that do achieve zero dissipation, Thms.
\ref{the:quant_retro} and \ref{the:quant_pre} imply that only \emph{certain}
processes have zero-dissipation quantum generators and, moreover, those
particular processes achieve no memory compression. The memory states, being
orthogonally encoded, take no advantage of the quantum setting to reduce their
memory cost.

\section{Concluding Remarks}

We identified the conditions under which local operations circumvent the
thermodynamic dissipation $\Delta S_{\mathrm{loc}}$ that arises from destroying
correlation. We started by showing how a useful theorem can be derived using
recent results on the fixed points of quantum channels. We applied it to the
setting of local operations to determine the necessary and sufficient
conditions for vanishing $\Delta S_{\mathrm{loc}}$ in classical and quantum
settings, with the aid of a generalized notion of quantum sufficient statistic.
We employed this fundamental result to review and extend previous results on
the thermodynamic efficiency of generators of stochastic processes. We
confirmed a recent conjecture regarding the conditions for vanishing $\Delta
S_{\mathrm{loc}}$ in a classical generator. And, then, we showed the exact same
conditions hold for quantum generators, even to the point of requiring
orthogonal encoding of memory states. This implies the profound result that
quantum memory compression and perfect efficiency ($\Delta S_{\mathrm{loc}}=0$)
are incompatible.

It is appropriate here to recall the lecture by Feynman in the early days of
thinking about quantum computing, in which he observed that quantum systems can
only be simulated on classical (even probabilistic) computers with great
difficulty, but on a fundamentally-quantum computer they could be more
realistically simulated \cite{Feyn82a}. Here, we considered the task of
simulating a classical stochastic process by two means: one by using
fundamentally-classical but probabilistic machines and the other by using a
fundamentally-quantum machine. Previous results generally indicated quantum
machines are advantageous in memory for this task, in comparison to their
classical counterparts. Historically, this led to a much stronger notion of
``quantum supremacy'' than Feynman proposed: quantum computers may be advantageous in \emph{all} tasks \cite{Pres12a}.

However, the quantum implementation we examined, though advantageous in memory,
requires nonzero dissipation in order to cash in on that advantage. Furthermore,
not every process necessarily has a quantum generator that achieves zero
dissipation. This is in sharp contrast to the classical outcome. And so, this
returns us to the spirit of Feynman's vision for simulating physics, in which it
may sometimes be the case that the best machine to simulate a classical
stochastic process is a classical stochastic computer---at least,
thermodynamically speaking.

To further exercise these results, further extensions must be made to
quantum generators, beyond the $q$-machine and its time reverse. We must determine if the exclusive relationship between
compression and zero dissipation continues to hold in such extensions. We pursue
this question in forthcoming work.

\section*{Acknowledgments}
\label{sec:acknowledgments}

The authors thank Fabio Anza, David Gier, Richard Feynman, Ryan James,
Alexandra Jurgens, Gregory Wimsatt, and Ariadna Venegas-Li for helpful
discussions and the Telluride Science Research Center for hospitality during
visits. As a faculty member, JPC similarly thanks the Santa Fe Institute. This
material is based upon work supported by, or in part by FQXi Grant
FQXi-RFP-IPW-1902, the U.S. Army Research Laboratory and the U. S. Army
Research Office under contract W911NF-13-1-0390 and grant W911NF-18-1-0028, and
via Intel Corporation support of CSC as an Intel Parallel Computing Center.

\appendix
\onecolumngrid
\clearpage
\begin{center}
\large{Supplementary Materials}\\
\vspace{0.1in}
\emph{\ourTitle}\\
\vspace{0.1in}
{\small
Samuel P. Loomis and James P. Crutchfield
}
\end{center}

\setcounter{equation}{0}
\setcounter{figure}{0}
\setcounter{table}{0}
\setcounter{page}{1}
\makeatletter
\renewcommand{\theequation}{S\arabic{equation}}
\renewcommand{\thefigure}{S\arabic{figure}}
\renewcommand{\thetable}{S\arabic{table}}

The Supplementary Materials give a quick notational overview, review (i) fixed
points of quantum channels, reversible computation, and sufficient statistics,
(ii) quantum implementations of classical generators and $q$-machines and their
thermodynamic costs, and (iii) provide details on example calculations. The
intention is that the main development be accessible while, together with the
Supplementary Materials, the full treatment becomes self-contained.

\section{Notation and Basic Concepts}

\subsection{State, Measurement, and Channels}

\emph{Quantum systems} are denoted by letters $A$, $B$, and $C$ and represented
by Hilbert spaces $\mathcal{H}_A$, $\mathcal{H}_B$, $\mathcal{H}_C$,
respectively. A system (say $A$) has state $\rho_A$ that is a positive bounded
operator on $\mathcal{H}_A$. The set of bounded operators on $\mathcal{H}_A$ is
denoted by $\mathcal{B}\left(\mathcal{H}_A\right)$ and the set of positive
bounded operators by $\mathcal{B}_{+}\left(\mathcal{H}_A\right)$.

Measurements $W$, $X$, $Y$, and $Z$ take values in the countable sets
$\mathcal{W}$, $\mathcal{X}$, $\mathcal{Y}$, and $\mathcal{Z}$, respectively. A
\emph{measurement} $X$ on system $A$ is defined by a set of Kraus operators
$\left\{K^{(x)}:x\in\mathcal{X}\right\}$ over $\mathcal{H}_A$ that satisfy the
completeness relation: $\sum_x K^{(x)\dagger}K^{(x)} = I_A$---the identity on
$\mathcal{H}_A$. Given state $\rho_A$ and measurement $X$, the probability of
outcome $x \in \mathcal{X}$ is:
\begin{align*}
  \Pr\left(X=x;\rho_A\right)
  := \mathrm{Tr}\left(K^{(x)} \rho_A K^{(x)\dagger}\right)
\end{align*}
and the state is transformed into:
\begin{align*}
\left.\rho_A\right|_{X=x} := \frac{K^{(x)} \rho_A K^{(x)\dagger}}{\Pr\left(x;\rho_A\right)}
  ~.
\end{align*}
If the Kraus operators $\left\{K^{(x)}:x\in\mathcal{X}\right\}$ are orthogonal
to one another and projective, then $X$ is called a \emph{projective
measurement}. Otherwise, $X$ is called a \emph{positive-operator valued
measure} (POVM).

Given a set of projectors $\{\Pi^{(x)}\}$ on a Hilbert space $\mathcal{H}_A$,
such that $\sum_x \Pi^{(x)} = I_A$ is the identity, we may decompose
$\mathcal{H}_A=\bigoplus_x \mathcal{H}_A^{(x)}$ into a direct sum of subspaces
$\mathcal{H}_A^{(x)}$, each one corresponding to the support of projector
$\Pi^{(x)}$. We will use the direct sum symbol $\bigoplus$ to indicate sums of
orthogonal objects. For instance, if $\rho^{(x)}_A$ is a state on
$\mathcal{H}_A^{(x)}$ for each $x$, then $\bigoplus_x \rho^{(x)}_A$ represents
a block-diagonal density matrix over $\mathcal{H}_A$. Similarly, if
$K^{(y|x)}_A$ represents a linear operator from $\mathcal{H}_A^{(x)}$ to
$\mathcal{H}_A^{(y)}$, then $\bigoplus_{x,y}K^{(y|x)}_A$ represents a linear
operator from $\mathcal{H}_A$ to itself, defined by a block matrix with
diagonal ($x=y$) and off-diagonal elements.

If $\mathcal{H}_A=\bigoplus_x \mathcal{H}_A^{(x)}$ and all the
$\mathcal{H}_A^{(x)}$ are unitarily equivalent to one another, then we can
equivalently write $\mathcal{H}_A= \mathcal{H}_{A_1}\otimes \mathcal{H}_{A_2}$
where $\mathcal{H}_{A_1}$ is equivalent to each of the $\mathcal{H}_A^{(x)}$
and $\mathcal{H}_{A_2}$ is equivalent to $\mathbb{C}^{|\mathcal{X}|}$.

Let $\rho_A$ and $\sigma_A$ be states in
$\mathcal{B}_{+}\left(\mathcal{H}_A\right)$. If, for every measurement $X$ and
outcome $x\in\mathcal{X}$, $\Pr\left(x;\sigma_A\right)>0$ implies that
$\Pr\left(x;\rho_A\right)>0$, then we write $\rho_A \ll \sigma_A$, indicating
that $\rho_A$ is \emph{absolutely continuous} with respect to $\sigma_A$.

To discuss classical systems, we eschew states and instead focus directly on
the measurements. In this setting, measurements are called \emph{random
variables}. Here, a random variable $X$ is defined over a countable set
$\mathcal{X}$, taking values $x\in\mathcal{X}$ with a specified probability
distribution $\Pr\left(X=x\right)$. When the variable can be inferred from
context, we will simply write the probabilities as $\Pr\left(x\right)$.

A random variable $X$ can generate a quantum state $\rho_{A}$ through an
\emph{ensemble} $\left\{\left(\Pr(X=x),\rho^{(x)}_A\right)\right\}$ of
potentially nonorthogonal states such that:
\begin{align*}
\rho_A = \sum_{x\in\mathcal{X}} \Pr\left(X=x\right) \rho^{(x)}_A
  ~.
\end{align*}

A \emph{quantum channel} is a linear map
$\mathcal{E}:\mathcal{B}\left(\mathcal{H}_A\right)\rightarrow
\mathcal{B}\left(\mathcal{H}_C\right)$ that is trace-preserving and completely
positive. These conditions are equivalent to requiring that there is a random
variable $X$ and a set of Kraus operators $\{K^{(x)}:x\in\mathcal{X}\}$ such
that:
\begin{align*}
  \mathcal{E}\left(\rho_A\right) = \sum_x K^{(x)} \rho_A K^{(x)\dagger}
  ~,
\end{align*}
for all $\rho_A$.\footnote{It is sufficient for our purposes to consider the
case where Kraus operators form a countable set. This holds, for instance, as
long as we work with separable Hilbert spaces.} To every channel $\mathcal{E}$
there is an \emph{adjoint} $\mathcal{E}^\dagger$:
\begin{align*}
\mathrm{Tr}\left(\mathcal{E}(\rho_A) M\right)
  = \mathrm{Tr}\left(\rho_A \mathcal{E}^\dagger(M)\right)
  ~,
\end{align*}
for every state $\rho_A\in \mathcal{B}_{+}\left(\mathcal{H}_A\right)$ and
operator $M\in \mathcal{B}\left(\mathcal{H}_A\right)$. The adjoint has the
form:
\begin{align*}
  \mathcal{E}^\dagger\left(M\right) = \sum_x K^{(x)\dagger} M K^{(x)}
  ~.
\end{align*}
Given a subspace $\mathcal{H}_B\subseteq \mathcal{H}_A$, we denote the
restriction of $\mathcal{E}$ to $\mathcal{H}_B$ as
$\left.\mathcal{E}\right|_{\mathcal{H}_B}:\mathcal{B}\left(\mathcal{H}_B\right)\rightarrow
\mathcal{B}\left(\mathcal{H}_C\right)$. A map
$\mathcal{E}:\mathcal{B}\left(\mathcal{H}_A\right)\rightarrow
\mathcal{B}\left(\mathcal{H}_A\right)$ is {\em ergodic} on a space
$\mathcal{H}_A$ if there is no proper subspace $\mathcal{H}_B\subseteq
\mathcal{H}_A$ such that the range of
$\left.\mathcal{E}\right|_{\mathcal{H}_B}$ is limited to
$\mathcal{B}\left(\mathcal{H}_B\right)$. $\mathcal{E}$ is ergodic if and only
if it has a unique stationary state $\mathcal{E}\left(\pi_A\right)=\pi_A$
\cite[and references therein]{Carb16a}.

The classical equivalent of a quantum channel is, of course, the
\emph{classical channel} that maps a set $\mathcal{X}$ to $\mathcal{Y}$
according to the conditional probabilities $\Pr\left(Y=y|X=x\right)$. This may
be alternately represented by the stochastic matrix $\mathbf{T}:=(T_{yx})$ such
that $T_{yx}=\Pr\left(Y=y|X=x\right)$. Stochastic matrices are also defined by
the conditions that $T_{yx}\geq 0$ for all $x\in\mathcal{X}$ and $y\in\mathcal{Y}$
and $\sum_{y} T_{yx} = 1$ for all $x\in\mathcal{X}$. A channel from
$\mathcal{X}$ to itself is \emph{ergodic} if there is no proper subset
$\mathcal{Y}\subset \mathcal{X}$ such that, for $x\in\mathcal{Y}$,
$\Pr\left(X'=x'|X=x\right)>0$ only when $x'\in\mathcal{Y}$ as well. This is
equivalent to requiring that $T_{x'x}:=\Pr\left(X'=x'|X=x\right)$ is an
irreducible matrix.

A classical channel $\Pr(Y=y|X=x)$ can be induced from a joint distribution
$\Pr(x,y)$ by the Bayes' rule $\Pr(y|x) := \Pr(x,y)/\sum_y \Pr(x,y)$, so that
the joint distribution may be written as $\Pr(x,y)=\Pr(x)\Pr(y|x)$. Given three
random variables $XYZ$, we write $X-Y-Z$ to denote that $X$, $Y$, and $Z$ form
a \emph{Markov chain}: $\Pr(x,y,z)=\Pr(x)\Pr(y|x)\Pr(z|y)$. The definition is
symmetric in that it also implies $\Pr(x,y,z)=\Pr(z)\Pr(y|z)\Pr(x|y)$.

\subsection{Entropy, Information, and Divergence}

The uncertainty of a random variable or measurement is considered to be a proxy
for its information content. Often, uncertainty is measured by the
\emph{Shannon entropy}:
\begin{align*}
H\left(X\right) := -\sum_{x\in\mathcal{X}} \Pr\left(x\right)\log_2 \Pr\left(x\right)
  ~,
\end{align*}
when it does not diverge. Given a system $A$ with state $\rho_A$, the
quantum uncertainty is the \emph{von Neumann entropy}:
\begin{align*}
H\left(A\right) := -\mathrm{Tr}\left(\rho_A \log_2 \rho_A\right)
  ~.
\end{align*}
When the same system may have many states in the given context, this is written
directly as $H\left(\rho_A\right)$. It is the smallest entropy that can be
achieved by taking a projective measurement of system $A$. The minimum is
attained by the measurement that diagonalizes $\rho_A$.

There are many ways of quantifying the information shared between two random
variables $X$ and $Y$. The most familiar is the \emph{mutual information}:
\begin{align*}
I\left(X:Y\right) & := H\left(X\right)+H\left(Y\right) - H\left(XY\right) \\
  & = \sum_{x,y} \Pr\left(x,y\right) \log_2
  \left(\frac{\Pr\left(x,y\right)}{\Pr\left(x\right)\Pr\left(y\right)}\right)
  ~.
\end{align*}
Corresponding to the mutual information are the \emph{conditional entropies}:
\begin{align*}
H\left(X\middle|Y\right) & := H\left(X\right) - I\left(X:Y\right) ~\text{and} \\
~\ H\left(Y\middle|X\right) & := H\left(Y\right) - I\left(X:Y\right)
 ~. 
\end{align*}
We will also have occasion to use the \emph{conditional mutual information} for three variables:
\begin{align*}
  I(X:Y|Z) := \sum_{x,y,z} \Pr(x,y,z)\log_2\left(\frac{\Pr(x,y|z)}{\Pr(x|z)\Pr(y|z)}\right)
  ~.
\end{align*}
Note that an equivalent condition for $X-Y-Z$
is the vanishing of the conditional mutual information: $I(X:Z|Y)=0$.

For a bipartite quantum state $\rho_{AB}$ the mutual information is usually taken to be the analogous quantity $I\left(A:B\right) = H(A) + H(B) - H(AB)$
and the conditional entropies as $H\left(A\middle|B\right) = H(AB)-H(B)$ and
so on.

We need a way to compare quantum systems and classical random variables.
Consider an ensemble $\left\{\left(\Pr(X=x),\rho^{(x)}_A\right)\right\}$
generating average state $\rho_A$ and define its \emph{Holevo quantity}:
\begin{align*}
I\left(A:X\right) := H\left(\rho_A\right) - \sum_{x} \Pr\left(x\right)H\left(\rho^{(x)}_A\right)
  ~.
\end{align*}
Consider the ensemble-induced state $\rho_{AB}$ defined by:
\begin{align*}
  \rho_{AB} = \sum_{x\in\mathcal{X}} \Pr\left(X=x\right) \rho_A^{(x)}\otimes \ketbra{x}
  ~,
\end{align*}
where $\left\{\ket{x}\right\}$ is an orthogonal basis on $\mathcal{H}_B$. Such
a state is called \emph{quantum-classical}, or \emph{classical-quantum} if the
role of $A$ and $B$ are swapped. In the first case, $I(A:B)=I\left(A:X\right)$,
where $I(A:B)$ is the mutual information of $\rho_{AB}$ and $I(A:X)$ is the
Holevo quantity of the ensemble.

Consider random variables $X_1$ and $X_2$ over the same set $\mathcal{X}$ 
with two possible distributions $\Pr(X_1=x)$ and $\Pr(X_2=x)$, respectively,
and suppose that whenever $\Pr(X_2=x)>0$ we have $\Pr(X_1=x)>0$ as well.
Then, we can quantify the difference between the two distributions with the
\emph{relative entropy}:
\begin{align*}
D\left(\Pr(X_1)\middle\| \Pr(X_2)\right) := \sum_{x\in\mathcal{X}} \Pr(X_1=x)\log_2 \frac{\Pr(X_1=x)}{\Pr(X_2=x)}
  ~.
\end{align*}
Similarly, two states $\rho_A$ and $\sigma_A$ with $\rho_A\ll \sigma_A$ may be
compared via the \emph{quantum relative entropy}:
\begin{align*}
  D\left(\rho_A\middle\| \sigma_A\right) := \mathrm{Tr}\left(\rho_A\log_2 \rho_A - \rho_A \log_2 \sigma_A\right)
  ~.
\end{align*}
The relative entropy leads to allied information-theoretic quantities. For
instance:
\begin{align*}
I(X:Y) =
D\left(\Pr\left(X,Y\right)\middle\|\Pr\left(X\right)\Pr\left(Y\right)\right)
  ~,
\end{align*}
and:
\begin{align*}
I(A:B) = D\left(\rho_{AB}\middle\|\rho_A\otimes \rho_B\right)
  ~.
\end{align*}

One of the most fundamental information-theoretic inequalities is the
monotonicity of the relative entropy under transformations. This is the
\emph{data processing inequality} \cite{Cove06a}. For a quantum channel
$\mathcal{E}$ it says:
\begin{align}
  D\left(\mathcal{E}\left(\rho_A\right)\middle\| \mathcal{E}\left(\rho_A\right)\right)
  \leq D\left(\rho_A\middle\| \sigma_A\right)
\label{eq:dpi}
\end{align}
The condition for equality requires constructing the \emph{Petz recovery
channel}:
\begin{align}
\mathcal{R}_\sigma(\cdot)
  = \sigma^{1/2}\mathcal{E}^\dagger
  \left(\mathcal{E}(\sigma_A)^{-1/2}\cdot\mathcal{E}(\sigma_A)^{-1/2}\right)\sigma^{1/2}
  ~.
\end{align}
It is easy to check that
$\mathcal{R}_\sigma\circ\mathcal{E}\left(\sigma_A\right) = \sigma_A$. A
markedly useful result is that $D\left(\mathcal{E}\left(\rho_A\right)\middle\|
\mathcal{E}\left(\rho_A\right)\right) = D\left(\rho_A\middle\| \sigma_A\right)$
if and only if $\mathcal{R}_\sigma\circ\mathcal{E}\left(\rho_A\right) = \rho_A$
as well \cite{Petz86a,Petz88a}.

Two other forms of the data processing inequality are useful to note here. The
first uses another quantity for measuring distance between states called the
\emph{fidelity}:
\begin{align}
  F\left(\rho,\sigma\right) := \left(\mathrm{Tr}\left[\sqrt{\sqrt{\rho}\sigma\sqrt{\rho}}\right]\right)^2
  ~.
\label{eq:fidelity}
\end{align} 
It takes value $F=0$ when the states $\rho$ and $\sigma$ are completely orthogonal
and value $F=1$ if and only if $\rho = \sigma$. The data processing inequality
for fidelity states that for any quantum channel $\mathcal{E}$:
\begin{align}
F\left(\mathcal{E}(\rho),\mathcal{E}(\sigma)\right)
  \geq F\left(\rho,\sigma\right)
  ~.
\label{eq:dpi-fid}
\end{align}
This is yet another way of saying that states map closer together under
a quantum channel $\mathcal{E}$.

The second form arises from applying \cref{eq:dpi} to the mutual information.
Let $\mathcal{E}_A:\mathcal{B}\left(\mathcal{H}_A\right)\rightarrow
\mathcal{B}\left(\mathcal{H}_C\right)$ be a quantum channel. The local
operation $\mathcal{E}_A\otimes I_B$ maps a bipartite system $AB$ to $CB$. The
data processing inequality \cref{eq:dpi} implies that we have $I(C:B)\leq
I(A:B)$.

In these terms, our thermodynamic efficiency goal---$\Delta S_{\mathrm{loc}} =
0$---translates into determining conditions for
equality---$I(C:B)= I(A:B)$---using the Petz recovery channel and channel fixed
points.

\begin{figure}[t]
  \centering
  \includegraphics[width=0.7\columnwidth]{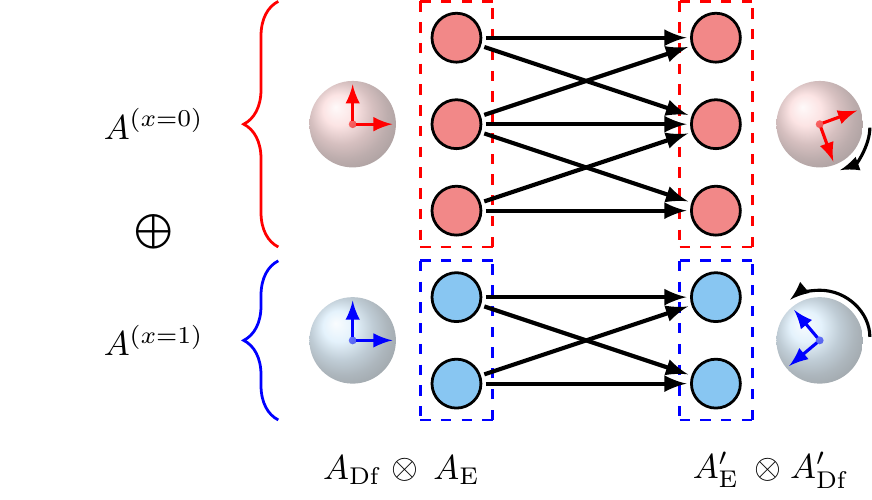}
\caption{{\bf Quantum channel decompositions}: Conserved measurement $X$
	divides the Hilbert space ``vertically'' via an orthogonal decomposition,
	$\mathcal{H}^{(x=0)}_A\oplus \mathcal{H}^{(x=1)}_A$, represented above by
	labels $A^{(x=0)}$ and $A^{(x=1)}$. For each value of $x$, there is a
	``horizontal'' decomposition into the tensor product of an ergodic subspace
	and a decoherence-free subspace: $\mathcal{H}^{(x)}_A =
	\mathcal{H}^{(x)}_{A_\rmE}\otimes \mathcal{H}^{(x)}_{A_\rmD}$, represented
	respectively by the labels $A_{\rmE}$ and $A_{\rmD}$. According to Theorem
	\ref{the:reverse}, information-theoretic reversibility requires storing
	data in the conserved measurement and the decoherence-free subspace. Any
	information stored coherently with respect to the conserved measurement
	(stored in the ergodic subspace) will be irreversibly modified under the
	channel's action.
  }
\label{fig:Channel}
\end{figure}

\section{Fixed Points and Reversible Computation}

To understand our result on local channels, an illustrative starting point is a
key result on fixed points of quantum channels \cite{Baum12a,Carb16a} that
leads to a natural decomposition, as illustrated in \cref{fig:Channel}.

\begin{The}[Channel and Stationary State Decomposition]
\label{the:fixedpoint}
Suppose $\mathcal{E}:\mathcal{B}\left(\mathcal{H}_A\right) \rightarrow
\mathcal{B}\left(\mathcal{H}_A\right)$ is a quantum channel, Hilbert space
$\mathcal{H}_A$ has a transient subspace $\mathcal{H}_{\mathrm{T}}$, and there
is a projective measurement $X=\left\{\Pi^{(x)}\right\}$ on
$\mathcal{H}_{\mathrm{T}}^{\perp}$ with countable outcomes $\mathcal{X}$, such
that $\mathcal{H}_A = \mathcal{H}_{\mathrm{T}} \oplus \left(\bigoplus_{x}
\mathcal{H}^{(x)}_A\right)$, where $\mathcal{H}^{(x)}_A$ is the support of
$\Pi^{(x)}$. Then:
\begin{enumerate}
\setlength{\topsep}{-5pt}
\setlength{\itemsep}{-5pt}
\setlength{\parsep}{-5pt}
\item Subspace $\mathcal{H}^{(x)}_A$ is preserved by $\mathcal{E}$, in that
	for all $\rho\in\mathcal{B}\left(\mathcal{H}^{(x)}_A\right)$,
	we have $\mathcal{E}\left( \rho\right)\in
	\mathcal{B}\left(\mathcal{H}^{(x)}_A\right)$.
\item Subspace $\mathcal{H}^{(x)}_A$ further decomposes into an ergodic subspace
	$\mathcal{H}^{(x)}_{A_\rmE}$ and decoherence-free subspace
	$\mathcal{H}^{(x)}_{A_\rmD}$:
\begin{align*}
	\mathcal{H}^{(x)}_A
	= \mathcal{H}^{(x)}_{A_\rmE} \otimes \mathcal{H}^{(x)}_{A_\rmD}
	~,
\end{align*}
	such that the Kraus operators of
	$\left.\mathcal{E}\right|_{\mathcal{H}_{\mathrm{T}}^{\perp}}$ decompose as
	\cite{Guan18a}:
\begin{align}\label{eq:kraus_auto}
    K^{(\alpha)} = \bigoplus_{x\in\mathcal{X}} K^{(\alpha,x)}_{A_\rmE} \otimes U^{(x)}_{A_\rmD}
\end{align}
	and the map $\mathcal{E}_{A_\rmE}^{(x)}\left(\cdot\right) := \sum_\alpha
	K^{(\alpha,x)}_{A_\rmE}\cdot K^{(\alpha,x)\dagger}_{A_\rmE}$ has a unique
	invariant state $\pi^{(x)}_{A_\rmE}$.
\item Any subspace of $\mathcal{H}$ satisfying the above two properties is, in
	fact, $\mathcal{H}^{(x)}_A$ for some $x\in\mathcal{X}$.
\end{enumerate}
Furthermore, if $\rho_A$ is any invariant state---that is,
$\rho_A = \mathcal{E}_A\left(\rho_A\right)$---then it decomposes as:
\begin{align}
  \rho_A = \bigoplus_{x\in\mathcal{X}}
  \Pr\left(x\right) \pi^{(x)}_{A_\rmE}\otimes \rho^{(x)}_{A_\rmD}
\label{eq:state_constraint_invariant}
  ~,
\end{align}
for any distribution $\Pr\left(x\right)$ and state $\rho^{(x)}_{A_\rmD}$
satisfying $U^{(x)}_{A_\rmD}\rho^{(x)}_{A_\rmD}U^{(x)\dagger}_{A_\rmD} =
\rho^{(x)}_{A_\rmD}$.
\end{The}

\Cref{fig:Channel} gives the geometric structure implied by the theorem.  The
ergodic subspace of quantum channel $\mathcal{E}$ has two complementary
decompositions. First, there is an orthogonal decomposition
$\mathcal{H}_{\mathrm{T}}^{\perp} = \bigoplus_x \mathcal{H}^{(x)}_A$ induced by
a projective measurement $X$ whose values are conserved by $\mathcal{E}$'s
action. This conservation is decoherent: only states compatible with $X$ are
stationary under the action of $\mathcal{E}$. $X$ is called the \emph{conserved
measurement} of $\mathcal{E}$ \cite{Albe19a}. Then, each $\mathcal{H}^{(x)}_A$
has a tensor decomposition $\mathcal{H}^{(x)}_A=\mathcal{H}^{(x)}_{A_\rmE}
\otimes \mathcal{H}^{(x)}_{A_\rmD}$ into an \emph{ergodic} (E) and a
\emph{decoherence-free} (Df) part. The decoherence-free subspace
$\mathcal{H}^{(x)}_{A_\rmD}$ undergoes only a unitary transformation
\cite{Guan18a}. The ergodic part $\mathcal{H}^{(x)}_{A_\rmE}$ is irreducibly
mixed such that there is a single stationary state.

This result's contribution here is to aid in identifying when the
data-processing inequality saturates. That is, using Thm. \ref{the:fixedpoint}
and the Petz recovery channel, we derive necessary and sufficient constraints
on the structures of $\rho_A$, $\sigma_A$, and $\mathcal{E}_A$ for determining
when $D\left(\mathcal{E}\left(\rho_A\right)
\middle\|\mathcal{E}\left(\sigma_A\right)\right) =
D\left(\rho_A\middle\|\sigma_A\right)$.

To achieve this, we recall a previously known result \cite{Moso04a,Moso05a},
showing that it can be derived using only the Petz recovery map and Thm.
\ref{the:fixedpoint}. The immediate consequence is a novel proof.

\begin{The}[Reversible Information Processing]
\label{the:reverse}
Suppose for two states $\rho_A$ and $\sigma_A$ on a Hilbert space
$\mathcal{H}_A$ and a quantum channel
$\mathcal{E}:\mathcal{B}\left(\mathcal{H}_A\right) \rightarrow
\mathcal{B}\left(\mathcal{H}_A\right)$, we have:
\begin{align*}
D\left(\rho_A\middle\|\sigma_A\right)
  = D\left(\rho_C\middle\|\sigma_C\right) 
  ~,
\end{align*}
where $\rho_C = \mathcal{E}_A\left(\rho_A\right)$ and $\sigma_C =
\mathcal{E}_A\left(\sigma_A\right)$. Then there is a measurement $X$ on $A$
with countable outcomes $\mathcal{X}$ and orthogonal decompositions
$\mathcal{H}_\mathcal{E} = \bigoplus_{x}\mathcal{H}_{A}^{(x)}$ and
$\mathcal{H}_{C}^{(x)}$ such that:
\begin{enumerate}
\setlength{\topsep}{-5pt}
\setlength{\itemsep}{-5pt}
\setlength{\parsep}{-5pt}
\item For all $\rho \in \mathcal{B}\left(\mathcal{H}^{(x)}_A\right)$ and
	$\mathcal{E}\left( \rho\right) \in
	\mathcal{B}\left(\mathcal{H}^{(x)}_{C}\right)$, the mapping
	$\left.\mathcal{E}\right|_{\mathcal{H}^{(x)}_A}$ onto
	$\mathcal{B}\left(\mathcal{H}^{(x)}_{C}\right)$ is surjective.
\item Subspaces $\mathcal{H}^{(x)}_A$ further decompose into:
\begin{align*}
\mathcal{H}^{(x)}_A
	& = \mathcal{H}^{(x)}_{A_\rmE} \otimes \mathcal{H}^{(x)}_{A_\rmD}
	~\text{and} \\
\mathcal{H}^{(x)}_C
    & = \mathcal{H}^{(x)}_{C_\rmE} \otimes \mathcal{H}^{(x)}_{C_\rmD}
	~,
\end{align*}
so that $\mathcal{H}^{(x)}_{C_\rmE}$ and $\mathcal{H}^{(x)}_{A_\rmD}$
are unitarily equivalent and the Kraus operators decompose as:
\begin{align}
\label{eq:kraus_mid}
  L^{(\alpha)} = \bigoplus_{x\in\mathcal{X}}
  L_{A_\rmE}^{(\alpha,x)}\otimes U^{(x)}_{A_\rmD}
  ~.
\end{align}  
\end{enumerate}
Furthermore, states $\rho_A$ and $\sigma_A$ satisfy:
\begin{subequations}
\label{eq:state_constraint_in}
\begin{align}
\rho_A &= \sum_{x\in\mathcal{X}} \Pr\left(x;\rho\right)\pi_{A_\rmE}^{(x)}\otimes \rho^{(x)}_{A_\rmD}\\
\sigma_A &= \sum_{x\in\mathcal{X}} \Pr\left(x;\sigma\right)\pi_{A_\rmE}^{(x)}\otimes \sigma^{(x)}_{A_\rmD}
\end{align}
\end{subequations}
for some $\pi_{A_\rmE}^{(x)}$.
And, their images are:
\begin{subequations}\label{eq:state_constraint_out}
\begin{align}
\rho_C & = \sum_{x\in\mathcal{X}} \Pr\left(x;\rho\right)
	\pi_{C_\rmE}^{(x)}\otimes \rho^{(x)}_{C_\rmD} \\
\sigma_C & = \sum_{x\in\mathcal{X}} \Pr\left(x;\sigma\right)
	\pi_{C_\rmE}^{(x)}\otimes \sigma^{(x)}_{C_\rmD}
  ~,
\end{align}
\end{subequations}
where $\pi_{C_\rmE}^{(x)} = \mathcal{E}^{(x)}_{A_\rmE}
\left(\pi^{(x)}_{A_\rmE}\right)$, $\rho^{(x)}_{C_\rmD} =
U^{(x)}_{A_\rmD}\rho^{(x)}_{A_\rmD} U^{(x)\dagger}_{A_\rmD}$, and
$\sigma^{(x)}_{C_\rmD} =
U^{(x)}_{A_\rmD}\sigma^{(x)}_{A_\rmD}U^{(x)\dagger}_{A_\rmD}$.
\end{The}

We know that $\mathcal{N}_\sigma:= \mathcal{R}_\sigma \circ \mathcal{E}$ must
have both $\rho_A$ and $\sigma_A$ as stationary distributions. Let $X$ be the
conserved measurement of $\mathcal{N}_\sigma $. It induces the decompositions
$\mathcal{H}_A = \mathcal{H}_{\mathrm{T}} \oplus \left(\bigoplus_{x}
\mathcal{H}^{(x)}_A\right)$ and
$\mathcal{H}^{(x)}_A=\mathcal{H}^{(x)}_{A_\rmE}\otimes
\mathcal{H}^{(x)}_{A_\rmD}$, as well as the state decompositions
\cref{eq:state_constraint_in}.

Now, we leverage the fact that $\left.\mathcal{N}_\sigma
\right|_{\mathcal{H}_{\mathrm{T}}^\perp}$ has Kraus decomposition of the form
\cref{eq:kraus_auto}. The net effect is summed up by two constraints:
\begin{enumerate}
\setlength{\topsep}{-5pt}
\setlength{\itemsep}{-5pt}
\setlength{\parsep}{-5pt}
\item For each $\alpha$, $K^{(\alpha)}$ maps each $\mathcal{H}^{(x)}_A$ to
	itself.
\item Let $K^{(\alpha,x)}$ be the block of $K^{(\alpha)}$ restricted to
	$\mathcal{H}^{(x)}_A$. Let $M$ be a complete projective measurement on
	$\mathcal{H}^{(x)}_{A_\rmD}$ with basis $\{\ket{m}\}$ and define the
	transformed basis $\{\ket{\widetilde{m}}=U^{(x)}\ket{m}\}$. Now, let
	$\mathcal{H}^{(x,m)}_{A} =
	\left\{\ket{\psi}\otimes\ket{m}:\ket{\psi}\in\mathcal{H}^{(x)}_{A_\rmE}\right\}$
	and similarly for $\mathcal{H}^{(x,\widetilde{m})}_{A}$. Then, $K^{(z,x)}$ maps
	$\mathcal{H}^{(x,m)}_{A}$ to $\mathcal{H}^{(x,\widetilde{m})}_{A}$.
\emph{This holds for any measurement $M$.}
\end{enumerate}

Proving \cref{eq:kraus_mid} requires these two constraints. Each is a form of
distinguishability criterion for the total channel $\left.\mathcal{N}_\sigma
\right|_{\mathcal{H}_{\mathrm{T}}^\perp}$. Since $\mathcal{N}_\sigma$ can tell
certain orthogonal outcomes apart, so too must $\mathcal{E}$. Or else,
$\mathcal{R}_\sigma$ would ``pull apart'' nonorthogonal states. However, this
is impossible for a quantum channel. By formally applying this notion to
constraints 1 and 2 above, we recover \cref{eq:kraus_mid}.

Let $L^{(\alpha)}$ be the Kraus operators of
$\left.\mathcal{E}\right|_{\mathcal{H}_{\mathrm{T}}^\perp}$. Then $K^{(\alpha)}
= \sigma^{1/2}L^{(\alpha)\dagger} \mathcal{E}(\sigma)^{1/2}L^{(z)}$. Now, if
$L^{(\alpha)}$ did not map each $\mathcal{H}^{(x)}_A$ to some orthogonal
subspace $\mathcal{H}^{(x)}_C$, then for some distinct $x$ and $x'$ there would
be $\ket{\psi}\in \mathcal{H}^{(x)}_A$ and $\ket{\phi}\in \mathcal{H}^{(x')}_C$
such that
$F\left(\mathcal{E}(\ketbra{\psi}),\mathcal{E}(\ketbra{\phi})\right)>0$; recall
\cref{eq:fidelity} defines fidelity.  However, we must also have
$F\left(\mathcal{N}_\sigma(\ketbra{\psi}),\mathcal{N}_\sigma(\ketbra{\phi})\right)=0$,
which is impossible by \cref{eq:dpi-fid} since as applying $\mathcal{R}_\sigma$
cannot reduce fidelity.  So, it must be the case that $L^{(\alpha)}$ maps each
$\mathcal{H}^{(x)}_A$ to some orthogonal subspace $\mathcal{H}^{(x)}_C$. This
proves Claim 1 in Thm. \ref{the:reverse}.

Let $L^{(\alpha,x)}$ be the block of $L^{(z)}$ restricted to
$\mathcal{H}^{(x)}_A$. Further, let $M$ be a complete measurement on
$\mathcal{H}^{(x)}_{A_\rmD}$. Then
$\left.\mathcal{E}\right|_{\mathcal{H}_A^{(x)}}$ must map each
$\mathcal{H}^{(x,m)}_{A}$ onto orthogonal subspaces $\mathcal{H}^{(x,m)}_{C}$
of $\mathcal{H}^{(x)}_{C}$, lest
$\left.\mathcal{N}_\sigma\right|_{\mathcal{H}_A^{(x)}}$ could not map each
$\mathcal{H}^{(x,m)}_{A}$ to orthogonal spaces
$\mathcal{H}^{(x,\widetilde{m})}_{A}$. This follows from fidelity, as in the
previous paragraph.

Now, let $L^{(\alpha,x,m)}:\mathcal{H}_{A_\rmE}^{(x)} \rightarrow
 \mathcal{H}^{(x,m)}_{C}$, so that:
\begin{align*}
  L^{(\alpha,x)} = \bigoplus_{m} L^{(\alpha,x,m)}\otimes \bra{m}
  ~.
\end{align*}
Let $N$ be another complete measurement on $\mathcal{H}^{(x)}_{A_\rmD}$
such that $\ket{n} = \sum_m w_{m,n}\ket{m}$, with $w_{m,n}$ a unitary matrix.
And, let $n,n'\in\mathcal{N}$ be distinct. We have for any $\ket{\psi}$ that:
\begin{align*}
\mathcal{E}^{(x)} \left(\ketbra{\psi}\otimes\ketbra{n}\right)
  & = \sum_{\alpha} \bigoplus_{m,m'} w_{m',n}w^\ast_{m,n}
  L^{(\alpha,x,m)}\ketbra{\psi}L^{(\alpha,x,m')\dagger} ~\text{and}\\
\mathcal{E}^{(x)} \left(\ketbra{\psi}\otimes\ketbra{n'}\right)
  & = \sum_{\alpha} \bigoplus_{m,m'} w_{m',n'}w^\ast_{m,n'}
  L^{(\alpha,x,m)}\ketbra{\psi}L^{(\alpha,x,m')\dagger}
  ~.
\end{align*}
Now, it must be that $\mathcal{E}^{(x)} \left(\ketbra{\psi}\otimes\ketbra{n}\right)$ and $\mathcal{E}^{(x)} \left(\ketbra{\psi}\otimes\ketbra{n'}\right)$
are orthogonal. However:
\begin{align*}
\mathrm{Tr}
\left[\mathcal{E}^{(x)}\left(\ketbra{\psi}\otimes\ketbra{n}\right)
  \mathcal{E}^{(x)}\left(\ketbra{\psi}\otimes\ketbra{n'}\right)\right]
  = \sum_{\alpha,\alpha'} \sum_{m} w_{m,n}
  w_{m,n'}^\ast \braket{\psi |  L^{(\alpha,x,m)\dagger}
  L^{(\alpha,x,m)}| \psi}
  ~.
\end{align*}
This vanishes for arbitrary $N$ and $\ket{\psi}$ only if
$L^{(\alpha|x,m)\dagger} L^{(\alpha'|x,m)}$ is independent of $m$ for each
$\alpha$ and $\alpha'$. This implies that $L^{(\alpha|x,m)} =
W^{(m)}L^{(\alpha|x)}_{\rmE}$ for some unitary $W^{(m)}$.

All of which leads one to conclude that the $\mathcal{H}^{(x,m)}_{C}$ for each
$m$ must be unitarily equivalent. And so, the decomposition
$\mathcal{H}^{(x)}_C = \bigoplus_m \mathcal{H}^{(x,m)}_{C}$ instead becomes a
tensor product decomposition $\mathcal{H}^{(x)}_C =
\mathcal{H}^{(x)}_{C_\rmE}\otimes \mathcal{H}^{(x)}_{C_\rmD}$ and further that
$L^{(\alpha,x)} = L^{(\alpha,x)}_{A_\rmE}\otimes V_{A_\rmD}^{(x)}$.

Finally, the constraints \cref{eq:state_constraint_out} follow from the form of
$\mathcal{E}$ and \cref{eq:state_constraint_in}. $\square$

Theorem \ref{the:reverse}'s main implication is that, when a channel
$\mathcal{E}$ acts, information stored in the conserved measurement and in the
decoherence-free subspaces is recoverable. Two states that differ in terms of
the conserved measurement and the decoherence-free subspaces remain different
and do not grow more similar under $\mathcal{E}$'s action. Conversely,
information stored in measurements not compatible with the conserved
measurement or stored in the ergodic subspaces is irreversibly garbled by
$\mathcal{E}$.

The next section uses this decomposition to study how locally acting channels
impact correlations between subsystems. This directly drives the thermodynamic
efficiency of local operations. Namely, for thermodynamic efficiency
correlations must be stored specifically in the conserved measurement and
decoherence-free subspaces of the local channel.

\section{Quantum Sufficient Statistics and Reversible Local Operations}
Let's review the definition of sufficient statistic in the main body.
Let $\rho_{AB}$ be a bipartite quantum state. A \emph{maximal local commuting
measurement} (MLCM) of $A$ for $B$ is any local measurement $X$ on system $A$
such that:
\begin{align*}
  \rho_{AB} = \bigoplus_{x} \Pr(X=x) \rho^{(x)}_{AB}
  ~,
\end{align*}
where:
\begin{align*}
\Pr(X=x) = \mathrm{Tr}\left((\Pi^{(x)}_X\otimes I_B) \rho_{AB}\right)
\end{align*}
and:
\begin{align*}
\Pr(X=x)\rho^{(x)}_{AB}
  = (\Pi^{(x)}_X\otimes I_B) \rho_{AB}(\Pi^{(x)}_X\otimes I_B)
  ~,
\end{align*}
and any further local measurement $Y$ on $\rho^{(x)}_{AB}$ disturbs the state:
\begin{align*}
\rho^{(x)}_{AB}
  \neq \sum_y (\Pi^{(y)}_Y\otimes I_B)\rho^{(x)}_{AB}(\Pi^{(y)}_Y\otimes I_B)
  ~.
\end{align*}
We call the states $\left\{ \rho^{(x)}_{AB}\right\}$ \emph{quantum correlation atoms}.

\begin{Prop}[MLCM uniqueness]
  Given a state $\rho_{AB}$, there is a unique MLCM of $A$ for $B$.
  \end{Prop} 
  
  Suppose there were two distinct MLCMs, $X$ and $Y$. Then:
  \begin{align*}
  \rho_{AB}
    = \sum_{y} \Pr(X=x) (\Pi^{(y)}_Y\otimes I_B)\rho_{AB}
    (\Pi^{(y)}_Y\otimes I_B)
    ~.
  \end{align*}
  This can be written as:
  \begin{align*}
  \rho_{AB}
    = \bigoplus_{x} \sum_{y} \Pr(X=x)
    (\Pi^{(y)}_Y\otimes I_B)\rho^{(x)}_{AB}(\Pi^{(y)}_Y\otimes I_B)
    ~.
  \end{align*}
  However, this means for each $x$:
  \begin{align*}
  (\Pi^{(y)}_Y\otimes I_B)\rho^{(x)}_{AB}(\Pi^{(y)}_Y\otimes I_B)
    = \rho^{(x)}_{AB}
    ~.
  \end{align*}
  So, $X$ is not a MLCM, giving a contradiction. $\square$

It will be helpful in our study of quantum generators to have the following fact as well:
\begin{Prop}[MLCM for a classical-quantum state]
  \label{the:cq_mclm}
  Given a classical-quantum state:
  \begin{align*}
  \rho_{AB}:= \sum_{x} \Pr\left(x\right)\ketbra{x}\otimes \rho_B^{(x)}
    ~,
  \end{align*}
  the MLCM is the most refined measurement $\Theta$ such that:
  \begin{align*}
  \rho_B^{(x)} = \sum_\theta \Pi^{(\theta)}\rho_B^{(x)}\Pi^{(\theta)}
  \end{align*}
  for all $x$.
  \end{Prop}
  
  Given that $\Theta$ is a commuting local measurement, the question is whether
  it is maximal. If it is not maximal, though, there is a refinement $Y$ that is
  also a commuting local measurement. By $\Theta$'s definition, there is an $x$
  such that $\rho_B^{(x)} \neq \sum_y \Pi^{(y)}\rho_B^{(x)}\Pi^{(y)}$. This
  implies $\rho_{AB} \neq \sum_y
  (I\otimes\Pi^{(y)})\rho_{AB}(I\otimes\Pi^{(y)})$, contradicting the assumption
  that $Y$ is commuting local. $\square$

Now, let $\rho_{AB}$ be a
bipartite quantum state and let $X$ be the MLCM of $A$ for $B$. We define the
{\em correlation equivalence} relation $x\sim x'$ over values of $X$ where
$x\sim x'$ if and only if $\rho^{(x)}_{AB} = (U\otimes I_B)\rho^{(x')}_{AB}
(U^{\dagger}\otimes I_B)$ for a local unitary $U$. 
  
Finally, we define the \emph{Minimal Local Sufficient Statistic} (MLSS)
$[X]_{\sim}$ as the equivalence class $[x]_{\sim}:=\{x':x'\sim x\}$ generated
by the relation $\sim$ between correlation atoms. Thus, our notion of
sufficiency of $A$ for $B$ is to find the most informative local measurement
and then coarse-grain its outcomes by unitary equivalence over their
correlation atoms. The correlation atoms and the MLSS $[X]_{\sim}$ together
describe the correlation structure of the system $AB$.

Using this definition and Theorem \ref{the:reverse}
we can prove our result on reversible local operations.

\begin{The}[Reversible local operations]
Let $\rho_{AB}$ be a bipartite quantum state and let $\mathcal{E}_A\otimes I_B$
be a local operation with
$\mathcal{E}_A:\mathcal{B}(\mathcal{H}_A)\rightarrow\mathcal{B}(\mathcal{H}_C)$.
Suppose $X$ is the MLCM of $\rho_{AB}$ and $Y$, that of $\rho_{CB} =
\mathcal{E}_A\otimes I_B\left(\rho_{AB}\right)$. The decomposition into
correlation atoms is:
\begin{align}
\label{eq:decomp}
\rho_{AB} & = \bigoplus_{x} \Pr{}_{A}\left(x\right) \rho_{A B}^{(x)}
  ~\text{and}\\
\rho_{CB} & = \bigoplus_{y} \Pr{}_{C}\left(y\right) \rho_{C B}^{(y)}
  ~.
\end{align}
Then, $I\left(A:B\right) = I\left(C:B\right)$ if and only if $\mathcal{E}_A$ can
be expressed by Kraus operators of the form:
\begin{align}
\label{eq:kraus_decomp}
K^{(\alpha)}
  = \bigoplus_{x,y} e^{i \phi_{xy\alpha}}\sqrt{\Pr(y,\alpha|x)} U^{(y|x)}
  ~,
\end{align}
where $\phi_{xy\alpha}$ is any arbitrary phase and
$\Pr\left(y,\alpha\middle|x\right)$ is a stochastic channel that is nonzero
only when $\rho_{A B}^{(x)}$ and $\rho_{C B}^{(y)}$ are equivalent up to a
local unitary operation $U^{(y|x)}$ that maps $\mathcal{H}^{(x)}_A$ to
$\mathcal{H}^{(y)}_C$.
\end{The}

We can apply the Reversible Information Processing Theorem (Thm.
\ref{the:reverse}) from the previous section here. This demands that there be a
measurement $X$ and a decomposition of the Hilbert space
$\mathcal{H}_A=\mathcal{H}_{A_{\rmE}}\otimes \mathcal{H}_{A_\rmD}$ such that:
\begin{align*}
\rho_{AB} &= \sum_{x} \Pr{}_{A}
  \left(x\right) \rho^{(x)}_{(AB)_{\rmE}} \otimes \rho_{(AB)_{\rmD}}^{(x)} \\
  \rho_{A} \otimes \rho_B
  &= \sum_{x} \Pr{}_{A}\left(x\right)
  \rho^{(x)}_{(AB)_{\rmE}} \otimes
  \left(\rho_{A_{\rmD}}^{(x)}\otimes \rho_{B_{\rmD}}^{(x)}\right)
  ~,
\end{align*}
such that $\mathcal{E}_A \otimes I_B$ conserves measurement $X$ and acts
decoherently on $(AB)_{\rmE}$ and coherently on $(AB)_{\rmD}$. However, the
local nature of $\mathcal{E}_A \otimes I_B$ makes it clear we can simplify this
decomposition to:
\begin{align*}
\rho_{AB}
  & = \sum_{x} \Pr{}_{A}\left(x\right)
  \rho^{(x)}_{A_{\rmE}}\otimes \left(\rho_{A_{\rmD} B}^{(x)}\right) \\
\rho_{A}\otimes \rho_B
  &= \sum_{x} \Pr{}_A\left(x\right)
  \rho^{(x)}_{A_{\rmE}} \otimes
  \left(\rho_{A_{\rmD}}^{(x)}\otimes \rho_{B}\right)
  ~,
\end{align*}
where $\mathcal{E}_A$ conserves the \emph{local} measurement $X$ on $A$ and
acts decoherently on $A_{\rmE}$ and acts as a local unitary
$U_{A_{\rmD}}\otimes I_B$ on $A_{\rmD}B$.

Suppose now, however, that given $X$ the variable $Y_x$ is the diagonalizing
measurement of $\rho^{(x)}_{A_\rmE}$ and $Z_x$ is the MLCM of
$\rho^{(x)}_{A_\rmD B}$. The joint measurement $XY_X Z_X$---where $X$ is
measured first and then the other two measurements are determined with
knowledge of its outcome---is the MLCM of $\rho_{AB}$. Note that for any $x$
and $z$, the outcomes $(x,y,z)$ and $(x,y',z)$ are correlation equivalent:
measurement $Y_X$ is completely decoupled from system $B$. Then, the MLSS
$\Sigma:= [X Y_X Z_X]_B$ is simply a function of $X$ and $Z_X$.

Since $XZ_X$ is conserved by the action of $\mathcal{E}_A\otimes I$---where
$X$ is the conserved measurement, while $Z_X$ is preserved through the unitary
evolution---the MLSS $\Sigma$ must be preserved and each correlation atom
is transformed only by a local unitary. This results in the form
\cref{eq:kraus_decomp}.

This proves that $I\left(A:B\right) = I\left(C:B\right)$ implies \cref{eq:kraus_decomp}.
The converse is straightforward to check. Let $\Sigma=[X]_B$
and let $I\left(A:B\middle|\Sigma=\sigma\right)$ be the mutual information
of $\rho_{AB}^{(x)}$ for any $x\in \sigma$. (This is the same for all such $x$ by local
unitary equivalence.) Then:
\begin{align*}
I(A:B) = \sum_\sigma
  \Pr\left(\Sigma=\sigma\right)I\left(A:B\middle|\Sigma=\sigma\right)
  ~.
\end{align*}
Similarly, let $\Sigma'=[Y]_B$
and let $I\left(C:B\middle|\Sigma'=\sigma'\right)$ be the mutual information
of $\rho_{CB}^{(y)}$ for any $y\in \sigma'$; then:
\begin{align*}
I(C:B) = \sum_{\sigma'}
  \Pr\left(\Sigma'=\sigma'\right)I\left(C:B\middle|\Sigma'=\sigma'\right)
  ~.
\end{align*}
Since $\mathcal{E}$ isomorphically maps each $\sigma$ to a unique $\sigma'$, such that 
$I\left(A:B\middle|\Sigma=\sigma\right)=I\left(C:B\middle|\Sigma'=\sigma'\right)$
by unitary equivalence, we must have $I(A:B)=I(C:B)$. $\square$

\section{Classical Generators}

Recall the definitions from the main body.
A classical generator is the physical implementation
of a hidden Markov model (HMM) \cite{Uppe97a}
$\mathfrak{G}=(\mathcal{S},\mathcal{X},\{T_{s's}^{(x)}\})$, where (here)
$\mathcal{S}$ is countable, $\mathcal{X}$ is finite, and for each
$x\in\mathcal{X}$, $\mathbf{T}^{(x)}$ is a matrix with values given by a
stochastic channel from $\mathcal{S}$ to $\mathcal{S}\times \mathcal{Y}$,
${T}^{(x)}_{s's} := \Pr{}_{\mathfrak{G}}(s',x|s)$. We define generators to use
\emph{recurrent} HMMs, which means the total transition matrix $T_{s's}:=\sum_x
T^{(x)}_{s's}$ is irreducible. In this case, there is a unique stationary
distribution $\pi_{\mathfrak{G}}(s)$ over states $\mathcal{S}$ satisfying
$\pi_{\mathfrak{G}}(s)>0$, $\sum_s \pi_{\mathfrak{G}}(s)=1$, and
$\sum_{s}T_{s's}\pi_{\mathfrak{G}}(s) = \pi_{\mathfrak{G}}(s')$.

The following probability distributions are particularly relevant:
\begin{align*}
  \Pr{}_{\mathfrak{G}}\left(x_1\dots x_\ell\right)
  &:= \sum_{s_0\dots s_\ell\in\mathcal{S}^{\ell+1}} 
  T^{(x_\ell)}_{s_\ell s_{\ell-1}}\dots T^{(x_1)}_{s_1 s_0} \pi(s_0)
  ~.\\
  \Pr{}_{\mathfrak{G}}\left(x_1\dots x_{t}, s_t\right)
  &:= \sum_{s_0\dots s_{t-1}\in\mathcal{S}^{t}} 
  T^{(x_t)}_{s_t s_{t-1}}\dots T^{(x_1)}_{s_1 s_0} \pi_{\mathfrak{G}}(s_0)
  ~.
\end{align*}

Also, recall the definition of a mergeable partition and retrodictive
equivalence. Each partition $\mathcal{P}=\left\{ \mathcal{P}_\theta\right\}$ of
$\mathcal{S}$ has a corresponding merged generator
$\mathfrak{G}^{\mathcal{P}}=(\mathcal{P},\mathcal{X},\{\widetilde{T}_{\theta'\theta}^{(x)}\})$
with transition dynamics given by:
\begin{align*}
  \widetilde{T}_{\theta'\theta}^{(x)}:=
   \sum_{\substack{s'\in\mathcal{P}_{\theta'}\\ s\in\mathcal{P}_\theta}} 
  \Pr{}_{\mathfrak{G}}\left(s',x|s\right) \pi(s|\theta)
  ~,
\end{align*}
where $\pi_{\mathfrak{G}}(s|\theta) =
\pi_{\mathfrak{G}}(s)/\pi_{\mathfrak{G}^{\mathcal{P}}}(\theta)$ and
$\pi_{\mathfrak{G}^{\mathcal{P}}}(\theta) = \sum_{s\in\mathcal{P}_\theta}
\pi_{\mathfrak{G}}(s)$. A partition $\left\{ \mathcal{P}_\theta\right\}$ is
\emph{mergeable} if the merged generator generates the same process as the
original.

One example of a partition is the retrodictive equivalence class. Two states
$s,s'\in\mathcal{S}$ are considered equivalent $s\sim s'$ if
$\Pr{}_{\mathfrak{G}}\left(x_1\dots
x_t|s\right)=\Pr{}_{\mathfrak{G}}\left(x_1\dots x_t|s'\right)$ for all words
$x_1\dots x_t$. The equivalence class $[S_t]_\sim$ is the sufficient statistic
of $S_t$ for predicting the past symbols $X_1\dots X_t$. The set
$\mathcal{P}_\sim := \{[s]_\sim:s\in\mathcal{S}\}$ of equivalence classes is a
partition on $\mathcal{S}$ that we index by $\sigma$. 

\begin{Prop}
\label{the:mergeable}
Given a generator $(\mathcal{S},\mathcal{X},\{T_{s's}^{(x)}\})$, the partition
$\mathcal{P}:=\mathcal{P}_\sim$ induced by retrodictive equivalence is
mergeable.
\end{Prop}
  
To see this, we follow a proof by induction. We first suppose that the
distribution $\Pr_{\mathfrak{G}^{\mathcal{P}}}\left(x_1\dots
x_t,\sigma_t\right)$ is equal to the distribution:
\begin{align*}
\Pr{}_{\mathfrak{G}}\left(x_1\dots x_t,\sigma_t\right)
  := \sum_{s_t\in\mathcal{P}_{\sigma_t}}
  \Pr{}_{\mathfrak{G}}\left(x_1\dots x_t,s_t\right)
  ~,
\end{align*}
for some $t$. Then, noting that:
\begin{align*}
  \Pr {}_{\mathfrak{G}} \left(x_1\dots x_{t+1},s_{t+1}\right) 
     & = \sum_{s_{t}}
    \Pr{}_{\mathfrak{G}}\left(x_1\dots x_{t}|s_{t}\right)
    \pi_{\mathfrak{G}}(s_{t}) T^{(x_{t+1})}_{s_{t+1} s_{t}} \\
     & = \sum_{\substack{\sigma_{t} \\ s_{t}\in\mathcal{P}_{\sigma_{t}}}}
    \Pr{}_{\mathfrak{G}}
    \left(x_1\dots x_{t}|\sigma_{t}\right)
    \pi_{\mathfrak{G}}(\sigma_{t}) 
    \pi_{\mathfrak{G}}(s_{t}|\sigma_{t})T^{(x_{t+1})}_{s_{t+1} s_{t}}
    ~,
\end{align*}
we have:
\begin{align*}
\Pr{}_{\mathfrak{G}}\left(x_1\dots x_{t+1},\sigma_{t+1}\right) 
   & = \sum_{\substack{\sigma_{t} \\s_{t}\in\mathcal{P}_{\sigma_{t}}\\ s_{t+1}\in\mathcal{P}_{\sigma_{t+1}}}}
  \Pr{}_{\mathfrak{G}^\mathcal{P}}\left(x_1\dots x_{t-1},\sigma_{t-1}\right)
  \widetilde{T}^{(x_{t+1})}_{\sigma_{t+1} \sigma_{t}} \\
  & = \Pr{}_{\mathfrak{G}^\mathcal{P}}\left(x_1\dots x_{t+1},\sigma_{t+1}\right)
  ~.
\end{align*}
Now, for $t=1$, we have:
\begin{align*}
\Pr{}_{\mathfrak{G}^\mathcal{P}}\left(x_1,\sigma_1\right)
  & = \sum_{\sigma_0}\pi_{\mathfrak{G}^\mathcal{P}}(\sigma_0)
  \widetilde{T}^{(x_1)}_{\sigma_1\sigma_0} \\
  & = \sum_{\sigma_0}\sum_{\substack{s_0\in \mathcal{P}_{\sigma_0}\\s_1\in \mathcal{P}_{\sigma_1}}}
  \pi_{\mathfrak{G}^\mathcal{P}}(\sigma_0)
  \pi_{\mathfrak{G}}(s_0|\sigma_0)
  {T}^{(x_1)}_{s_1 s_0} \\
   & = \sum_{s_1} \Pr{}_{\mathfrak{G}}\left(x_1,s_1\right)
  ~.
\end{align*}
  
Then, by induction, $\Pr_{\mathfrak{G}^\mathcal{P}}\left(x_1\dots
x_t,\sigma_t\right)=\Pr_{\mathfrak{G}}\left(x_1\dots x_t,\sigma_t\right)$ for
all $t$. By summing over $\sigma_t$, we have
$\Pr_{\mathfrak{G}^\mathcal{P}}\left(x_1\dots
x_t\right)=\Pr_{\mathfrak{G}}\left(x_1\dots x_t\right)$ for all $t$ as well.
$\square$

Using this result and the Reversible Local Operations Theorem, we can establish
the following.

\begin{The}\label{the:class_retro}
A generator $\mathfrak{G}=(\mathcal{S},\mathcal{X},\{T_{s's}^{(x)}\})$
satisfies $I\left(S_t :X_{1}\dots X_{t} \right)= I\left(S_{t+1}
X_{t+1}:X_{1}\dots X_{t}\right)$ for all $t$ if and only if the retrodictively
state-merged generator
$\mathfrak{G}^{\mathcal{P}}=(\mathcal{P}_{\sim},\mathcal{X},\{\widetilde{T}_{\sigma'\sigma}^{(x)}\})$
satisfies $\widetilde{T}_{\sigma'\sigma}^{(x)}\propto
\delta_{\sigma,f(\sigma',x)}$ for some function
$f:\mathcal{S}\times\mathcal{X}\rightarrow \mathcal{S}$.
\end{The}

To see this, recall from Cor. \ref{the:reversible_local_class} that 
$I\left(S_t :X_{1}\dots X_{t} \right)= I\left(S_{t+1} X_{t+1}:X_{1}\dots
X_{t}\right)$ only if $\Pr{}_{\mathfrak{G}}\left(s_{t+1},x_{t+1}|s_t\right)>0$
implies:
\begin{align*}
\Pr{}_{\mathfrak{G}} \left(x_1\dots x_t |s_{t+1},x_{t+1} \right)
	= \Pr{}_{\mathfrak{G}}\left(x_1\dots x_t |s_{t}\right) 
  ~.
\end{align*}
Now, note that:
\begin{align*}
\Pr{}_{\mathfrak{G}}\left(x_1\dots x_{t+1}|s_{t+1} \right) 
  & = \Pr{}_{\mathfrak{G}}\left(x_1\dots x_t |s_{t+1},x_{t+1}\right)
  \Pr\left(x_{t+1}|s_{t+1} \right) \\
  &  = \Pr{}_{\mathfrak{G}}\left(x_1\dots x_t |s_t\right)
  \Pr\left(x_{t+1}|s_{t+1} \right) 
  ~.
\end{align*}
Rearranging and using the retrodictive equivalence partitions
$\sigma_t := [s_t]_\sim$ and $\sigma_{t+1}:=[s_{t+1}]_\sim$, we have:
\begin{align}
\label{eq:function_def}
  \Pr{}_{\mathfrak{G}^\mathcal{P}}\left(x_1\dots x_{t}|\sigma_{t} \right) = 
  \frac{\Pr{}_{\mathfrak{G}^\mathcal{P}}\left(x_1\dots x_{t+1}|\sigma_{t+1} \right) }
  {\Pr{}_{\mathfrak{G}^\mathcal{P}}\left(x_{t+1}|\sigma_{t+1} \right)}
  ~.
\end{align}

Define a function $f:\mathcal{S}\times\mathcal{X}\rightarrow \mathcal{S}$ as
follows. For a given $\sigma'$ and $x$, let $f(\sigma',x)$ be the equivalence
class such that:
\begin{align*}
\Pr{}_{\mathfrak{G}^\mathcal{P}}\left(\cdot|f(\sigma',x)\right)
  = \frac{\Pr{}_{\mathfrak{G}^\mathcal{P}}\left(\cdot, x|\sigma'\right) }
  {\Pr{}_{\mathfrak{G}^\mathcal{P}}\left(x|\sigma'\right)}
  ~.
\end{align*}
Such an equivalence class $f(\sigma',x)$ must exist by \cref{eq:function_def}.
It is unique since, by definition, equivalence classes $\sigma$ have unique
distributions $\Pr\left(\cdot|\sigma\right)$. Then $\sigma_{t} =
f(\sigma_{t+1},x_{t+1})$ is a requirement for
$\Pr{}_{\mathfrak{G}^\mathcal{P}}\left(\sigma_{t+1},x_{t+1}|\sigma_t\right) >
0$. If we then take the merged generator $\mathfrak{G}^\mathcal{P}$, we must
have $\widetilde{T}_{\sigma'\sigma}^{(x)}>0$ only when $\sigma = f(\sigma',x)$. 

Conversely, suppose that the retrodictively state-merged generator
$\mathfrak{G}^{\mathcal{P}}=(\mathcal{P}_{\sim},\mathcal{X},\{\widetilde{T}_{\sigma'\sigma}^{(x)}\})$
satisfies $\widetilde{T}_{\sigma'\sigma}^{(x)}\propto
\delta_{\sigma,f(\sigma',x)}$ for some function
$f:\mathcal{S}\times\mathcal{X}\rightarrow \mathcal{S}$.  Then, for a given
$s_t$, its equivalence class $[s_t]_{\sim}$ is always a function of the next
state and symbol: $[s_t]_{\sim}=f\left([s_{t+1}]_{\sim},x_{t+1}\right)$. This
implies the Markov chain $[S_t]_{\sim} - X_{t+1}S_{t+1}- X_0\dots X_{t}$.
However, we also have the chain $X_{t+1}S_{t+1} - [S_t]_{\sim} - X_0\dots
X_{t}$ from the basic Markov property of the generator. Therefore, we must
have:
\begin{align*}
I\left(S_{t+1} X_{t+1}:X_{1}\dots X_{t}\right)
   & = I\left([S_t]_{\sim} :X_{1}\dots X_{t} \right) \\
   & = I\left(S_t :X_{1}\dots X_{t} \right)
  ~.
\end{align*}
$\square$

Last, we connect this result to previous literature on the efficiency of
retrodictors by showing that our conditions for efficiency imply that the
generator is retrodictive. Recall that any generator whose state $S_t$ is a
sufficient statistic of $\overrightarrow{X}_t$ for $X_1\dots X_t$ is called a
\emph{retrodictor}.

\begin{Prop}
A generator $\mathfrak{G}=(\mathcal{S},\mathcal{X},\{T_{s's}^{(x)}\})$ whose
retrodictively state-merged generator
$\mathfrak{G}^{\mathcal{P}}=(\mathcal{P}_{\sim},\mathcal{X},\{\widetilde{T}_{\sigma'\sigma}^{(x)}\})$
satisfies $\widetilde{T}_{\sigma'\sigma}^{(x)}\propto
\delta_{\sigma,f(\sigma',x)}$ for some function
$f:\mathcal{S}\times\mathcal{X}\rightarrow \mathcal{S}$ is a retrodictor.
\end{Prop}

This follows since $\mathfrak{G}^{\mathcal{P}}$, being co-unifilar and already
retrodictively state-merged, must be the reverse \eM with states $\Sigma_t$.  It
is clear from the proof of Prop. \ref{the:mergeable} that for all $t$, we have
the Markov chain $(X_1\dots X_t) - \Sigma_t - S_t$. Since $\Sigma_t$ is also the
minimum sufficient statistic of $\overrightarrow{X}_t$ for $X_1 \dots X_t$, we
must have the Markov chain $(X_1\dots X_t) - \overrightarrow{X}_t - S_t$.
Combined with the Markov chain $(X_1\dots X_t) - S_t - \overrightarrow{X}_t$, we
conclude that $S_t$ is a sufficient statistic of $\overrightarrow{X}_t$ for
$X_1\dots X_t$. $\square$

\section{Forward and Reverse $q$-Machines}

\subsection{$q$-Machines and Their Time Reversals}

A $q$-machine is constructed from a forward \eM
$\mathfrak{G}=(\mathcal{S},\mathcal{X},\{T_{s's}^{(x)}\})$ by choosing any set
of phases $\{\phi_{xs}:x\in\mathcal{X},s\in\mathcal{S}\}$ and constructing an
encoding $\{\ket{\psi_s}:s\in\mathcal{S}\}$ of the memory states $\mathcal{S}$
into a Hilbert space $\mathcal{H}_S$, and a set of Kraus operators
$\{K^{(x)}:x\in\mathcal{X}\}$ on said Hilbert space. The phases, encoding
states and the Kraus operators satisfy the formula:
\begin{align}
K^{(x)}\ket{\psi_s}
  = e^{i\phi_{xs}}\sqrt{T^{(x)}_{f(s,x),s}}\ket{\psi_{f(s,x)}}
  ~.
\label{eq:qdyn}
\end{align}
This expression implicitly defines the Kraus operators given the encoding
$\{\ket{\psi_s}\}$. The encoding, in turn, is determined up to a unitary
transformation by the following constraint on their overlaps:
\begin{align}
\braket{\psi_r|\psi_s} = \sum_{x\in\mathcal{X}}
  e^{i(\phi_{xs}-\phi_{xr})}\sqrt{T^{(x)}_{r',r}T^{(x)}_{s',s}}
  \braket{\psi_{r'}|\psi_{s'}}
  ~,
\label{eq:overlaps}
\end{align}
where $r'=f(r,x)$ and $s'=f(s,x)$ \cite{Liu19a}.

Now, let $\Omega_{rs}:=\braket{\psi_r|\psi_s} $ be the unique solution to
\cref{eq:overlaps}, which is effectively an eigenvalue equation for a
superoperator. Notice that this is entirely determined by the phases
$\{\phi_{xs}\}$ and the dynamics $\{T_{s's}^{(x)}\}$ of the original \eM. It
can be computed without any reference to encoding states or Kraus operators.
Indeed, once $\Omega_{rs}$ is determined, the encoding states and Kraus
operators can be explicitly constructed.

Let $\sqrt{\pi_r \pi_s}\Omega_{rs} = \sum_\alpha U_{r\alpha} U_{s\alpha}^\ast
\lambda_\alpha$ be the singular value decomposition of $\sqrt{\pi_r
\pi_s}\Omega_{rs}$ into a unitary $U_{i\alpha}$ and singular values
$\lambda_\alpha>0$. Suppose $\alpha=1,\dots,d$. Then given any computational
basis $\{\ket{\alpha}:\alpha=1,\dots,d\}$, we can construct:
\begin{align}
\label{eq:states_eigen}
\ket{\psi_s} &= \sum_\alpha \sqrt{\frac{\lambda_\alpha}{\pi_s}}
U_{s\alpha}^\ast
  \ket{\alpha} ~\text{and}\\
  \label{eq:ops_eigen}
  K^{(x)} &= \sum_{\substack{\alpha,\beta,s}} e^{i\phi_{xs}} U_{s'\beta}^\ast U_{s\alpha}
  \sqrt{\frac{\lambda_{\beta}\pi_{s}}{\lambda_\alpha \pi_{s'}}T^{(x)}_{s's}}
  \ket{\beta}\bra{\alpha}
  ~.
\end{align}
It is easy to check that $\braket{\psi_r|\psi_s} =\Omega_{rs}$ and that
\cref{eq:qdyn} is satisfied by this construction. Notice that:
\begin{align*}
\rho_\pi = \sum_s \pi_s \ketbra{\psi_s} = \sum_\alpha \lambda_\alpha \ketbra{\alpha}
  ~.
\end{align*}
So the computational basis $\alpha$ is the diagonal basis of the stationary state $\rho_\pi$.

This explicit construction is useful for time-reversing the $q$-machine.
Let $\mathfrak{G}$ be a reverse \eM.
The time-reverse of a generator $\mathfrak{G}$ is the generator 
$\widetilde{\mathfrak{G}}=(\mathcal{S},\mathcal{X},\{\widetilde{T}{}_{s's}^{(x)}\})$
where $\widetilde{T}{}_{s's}^{(x)} = \pi_s T_{s's}^{(x)}/\pi_{s'}$.
$\widetilde{\mathfrak{G}}$ is the forward \eM of the reverse process.
From it, we can construct a $q$-machine, with Kraus operators expressed
by $\left\{\widetilde{K}{}^{(x)}\right\}$.
Recall that the generated process of the $q$-machine is given by
\begin{align*}
  \Pr{}_{\widetilde{\mathfrak{G}}}\left(x_1\dots x_t\right)
    := \mathrm{Tr}\left[\widetilde{K}{}^{(x_t\dots x_1)}\rho_\pi
	\widetilde{K}{}^{(x_t\dots x_1)\dagger}\right]
    ~.
  \end{align*}
This is the time-reverse of the process generated by $\mathfrak{G}$, expressed
in the equation $\Pr{}_{\widetilde{\mathfrak{G}}}\left(x_1\dots x_t\right)
=\Pr{}_{{\mathfrak{G}}}\left(x_t\dots x_1\right)$. In terms of the $q$-machine,
we can write:
\begin{align*}
\Pr{}_{{\mathfrak{G}}}\left(x_1\dots x_t\right)
  & := \mathrm{Tr}\left[\widetilde{K}{}^{(x_t)}\dots \widetilde{K}{}^{(x_1)}
  \rho_\pi \widetilde{K}{}^{(x_1)\dagger}\widetilde{K}{}^{(x_t)\dagger}\right] \\
  & = \mathrm{Tr}\left[{K}{}^{(x_t)}\dots {K}{}^{(x_1)}
  \rho_\pi {K}{}^{(x_1)\dagger}\dots {K}{}^{(x_t)\dagger}\right]
  ~,
\end{align*}
where ${K}{}^{(x)} =
\rho_\pi^{1/2}\widetilde{K}{}^{(x)\dagger}\rho_\pi^{-1/2}$. This is,
essentially, the Petz reversal of the POVM $\{\widetilde{K}{}^{(x)}\}$, and
it constitutes a formal time-reversal of the quantum process \cite{Croo08a}.

Computing $K^{(x)}$ is straightforward using \cref{eq:states_eigen}, as this
gives the Kraus operators in the diagonal basis of $\rho_\pi$, where it is
easiest to compute $\rho^{1/2}_\pi$ and its inverse. We have:
\begin{align*}
{K}{}^{(x)} = \sum_{\substack{\alpha,\beta,s}} e^{-i\phi_{xs}} U_{s'\beta}
  U_{s\alpha}^{\ast}\sqrt{\frac{\pi_{s}}{\pi_{s'}}\widetilde{T}^{(x)}_{s's}}
  \ket{\alpha}\bra{\beta}
  ~.
\end{align*}
Now, take the basis $\ket{{\psi}{}_{s}} = \sum_\alpha U_{s\alpha}^{\ast}
\ket{\alpha}$. In this basis: 
\begin{align*}
{K}{}^{(x)} = \sum_{\substack{s'}} e^{-i\phi_{xs'}} \sqrt{{T}^{(x)}_{s's}}
  \ket{{\psi}{}_{s'}}\bra{{\psi}{}_{s}}
  ~.
\end{align*}
Thus, we see that the basis $\{\ket{{\psi}{}_s}\}$ and Kraus operators
$\{{K}{}^{(x)} \}$ form a reverse $q$-machine as described in the main
body---one that compresses the reverse \eM ${\mathfrak{G}}$.

Note that the stationary state of a time-reversed $q$-machine is just the
stationary state of the original $q$-machine---this is not altered under time
reversal. However, we find a new expression for the stationary state, in terms
of the basis $\{\ket{{\psi}{}_s}\}$:
\begin{align*}
\rho_\pi & = \sum_{r,s,\alpha}\lambda_\alpha
  U_{r\alpha}^\ast U_{s\alpha}\ket{{\psi}{}_s}\bra{{\psi}{}_r} \\
  & = \sum_{r,s} \sqrt{\pi_r \pi_s} \Omega_{sr}
  \ket{{\psi}{}_s}\bra{{\psi}{}_r}
  ~.
\end{align*}
So, $\rho_\pi$ is generally not diagonal in the basis $\{\ket{{\psi}{}_s}\}$.
The extent to which $\rho$ commutes with $\{\ket{{\psi}{}_s}\}$ is the extent
to which $\Omega_{rs}$ is block-diagonal. 

\subsection{Efficiency of $q$-Machines}

To establish our first theorem relating memory to efficiency we turn to forward
$q$-machines. First, we must prove a result regarding the synchronization of
$q$-machine states.

\begin{Prop}[Synchronization of $q$-machines]\label{prop:synch-qm}
Let $\rho_{x_1\dots x_t} := \Pr{}_{\mathfrak{G}}\left(x_1\dots x_t\right)^{-1}
K^{(x_t\dots x_1)}\rho_\pi K^{(x_t\dots x_1)\dagger}$ be the state of the
$q$-machine's memory system after seeing the word $x_1\dots x_t$. Further, let
$\hat{s}(x_1\dots x_t):= \mathrm{argmax}_{s_t} \Pr(s_t|x_1\dots x_t)$ be the
most likely memory state after seeing the word $x_1\dots x_t$ and let
$F(x_1\dots x_t):= F\left(\rho_{x_1\dots x_t},\ketbra{\psi_{\hat{s}}}\right)$
be the fidelity between the quantum state of the memory system and the most
likely encoded state. Then, there exist $0<\alpha<1$ and $K>0$ such that, for
all $t$:
\begin{align*}
\Pr\left(F(x_1\dots x_t)<1-\alpha^t\right) \leq K\alpha^t
  ~.
\end{align*}
This shows that the quantum state of the memory system converges to a single
encoded state with high probability.
\end{Prop}
  
This follows straightforwardly from a similar statement about \eMs
\cite{Trav10a,Trav10b,Trav12a}. Let $Q(x_1\dots x_t):= 1-\Pr(\hat{s}|x_1\dots
x_t)$ be the probability of \emph{not} being in the most likely state after word
$x_1\dots x_t$. Then there exist $0<\alpha<1$ and $K>0$ such that, for all $t$:
\begin{align*}
  \Pr\left(Q(x_1\dots x_t)>\alpha^t\right) \leq K\alpha^t
  ~.
\end{align*}
Note that, from the Kraus operator definition, the quantum state of the memory
system after word $x_1\dots x_t$ is:
\begin{align*}
\rho_{x_1\dots x_t}
  = \sum_{s_t}\Pr\left(s_t|x_1\dots x_t\right) \ketbra{\psi_{s_t}}
  ~.
\end{align*}
Now, the fidelity can be computed as:
\begin{align*}
F(x_1\dots x_t) & = \mathrm{Tr}
  \left[\sqrt{\ketbra{\psi_{\hat{s}}}
  \rho_{x_1\dots x_t}\ketbra{\psi_{\hat{s}}}}\right]^2 \\
  & = \braket{\psi_{\hat{s}} | \rho_{x_1\dots x_t} | \psi_{\hat{s}}} \\
  & = \Pr\left(\hat{s}|x_1\dots x_t\right) 
     + \sum_{s_t\neq \hat{s}}\Pr\left(s_t|x_1\dots x_t\right)
    \left|\braket{\psi_{s_t}|\psi_{\hat{s}}}\right|^2 \\
  & \geq 1 - Q\left(x_1\dots x_t\right)
  ~.
\end{align*}
And so, from the synchronization theorem for \eMs, we have:
\begin{align*}
  \Pr\left(F(x_1\dots x_t)<1-\alpha^t\right) \leq K\alpha^t
  ~.
\end{align*}
for all $t$. $\square$

The second notion we must introduce is a simple partition that may be
constructed on the memory states of the \eM for a given quantum implementation.
Specifically, we define the \emph{maximal commuting partition} on $\mathcal{S}$
to be the most refined partition $\{\mathcal{B}_\theta\}$ such that the overlap
matrix $\braket{\psi_r|\psi_s}$ is block-diagonal. That is,
$\{\mathcal{B}_\theta\}$ is such that $\braket{\psi_r|\psi_s}=0$ if $r\in
\mathcal{B}_\theta$ and $s\in\mathcal{B}_{\theta'}$ for $\theta\neq \theta'$.
From this partition we construct the maximally commuting local measurement
required to define sufficient statistics.

\begin{Prop}
Let $\rho_{\mathfrak{G}}(t)$ be the state of the system $A_1\dots A_t S_t$ at time $t$. Let $\Theta$ be the projective measurement on $\mathcal{H}_S$ corresponding to the MCP of the quantum generator. Then, for sufficiently large $t$, $\Theta$ is the MLCM of $S_t$. Similarly, $X_t \Theta$ is the MLCM of $A_t S_t$.
\end{Prop}
  
By Prop. \ref{the:cq_mclm}, the MLCM must leave each $\rho_{x_1\dots x_t}$
unchanged, for all $t$. This is true for $\Theta$. The question is if any
nontrivial refinement, say $Y$, of $\Theta$ can do so. Now, realize that for
any $\epsilon>0$, there is sufficiently large $t$, so that for each state $s$
there must be at least one word $x_1\dots x_t$ satisfying
$F\left(\rho_{x_1\dots x_t},\ketbra{\psi_s}\right)>1-\epsilon$. Then, for
sufficiently large $t$, it must be the case that there exists a word $x_1\dots
x_t$ such that $Y$ modifies $\rho_{x_1\dots x_t}$, because $Y$ (by virtue of
being a refinement of the maximal commuting partition) cannot commute with all
the $\ketbra{\psi_s}$. Therefore, $\Theta$ is the maximal commuting local
measurement. That $X_t \Theta$ is the MLCM of $A_t S_t$ follows from similar
considerations. $\square$

We can now prove the following.

\begin{The}[Maximally-efficient quantum generator]
Let $\mathfrak{G}=(\mathcal{S},\mathcal{X},\{T_{s's}^{(x)}\})$ be a given
process' \eM. Suppose we build from it a quantum generator with encoding
$\left\{\psi_s\right\}$ and Kraus operators $\{K^{(x)}\}$. Let
$\mathcal{B}:=\{\mathcal{B}_\theta\}$ be the MCP of $\mathcal{S}$. Then the
quantum generator has $\Delta S_{\mathrm{loc}}=0$ if and only if the partition
$\mathcal{B}$ is trivially maximal---in that $|\mathcal{B}_\theta|=1$ for each
$\theta$---and the retrodictively state-merged generator
$\mathfrak{G}^{\mathcal{B}}$ of $\mathfrak{G}$ is co-unifilar.
\end{The}

We define
$\rho_{\mathfrak{G}}^{(\theta)}(t):=\Pi^{(\theta)}\rho_{\mathfrak{G}}(t)\Pi^{(\theta)}$,
so that $\rho_{\mathfrak{G}}(t)=\bigoplus_\theta
\rho_{\mathfrak{G}}^{(\theta)}(t)$.  For the equivalence relation we say that
$\theta\sim\theta'$ if $\rho_{\mathfrak{G}}^{(\theta)}(t)$ and
$\rho_{\mathfrak{G}}^{(\theta')}(t)$ are unitarily equivalent on the subsystem
$S$. This defines a partition $\mathcal{P}=\left\{[\theta]_{\sim}\right\}$ over
the set $\{\mathcal{B}_\theta\}$.

The quantum channel:
\begin{align*}
  \mathcal{E}\left(\rho\right) := \sum_x K^{(x)}\rho K^{(x)\dagger}\otimes \ketbra{x}
\end{align*}
has Kraus operators $L^{(x)}:= K^{(x)}\otimes \ket{x}$. The MLCM of $A_t S_t$ is
$X_t \Theta_t$. Now, Thm. \ref{the:reversible_local} demands that the Kraus
operators $L^{(x)}$ have the form:
\begin{align*}
L^{(x)} & = \bigoplus_{\theta,\theta'}
  \sqrt{\Pr(\theta',x|\theta)} U^{(x, \theta'|\theta)} \\
  & = \bigoplus_{\theta,\theta'}
  \sqrt{\Pr(\theta',x|\theta)} \ket{x}\otimes U_{\theta\mapsto \theta'}^{(x)}
  ~.
\end{align*}
This implies that the quantum-generator Kraus operators have the form:
\begin{align}\label{sm:kraus_form}
K^{(x)} = \bigoplus_{\theta,\theta'}
  \sqrt{\Pr(\theta',x|\theta)} U^{(x)}_{\theta\mapsto \theta'}
  ~.
\end{align}
The values $\Pr(\theta',x|\theta)$ must be positive only when $\theta' x \sim
\theta$.

Now, we note that this means there is a machine
$(\mathcal{B},\mathcal{X},\{\hat{T}_{\theta'\theta}^{(x)}\})$ with transition
probabilities $\hat{T}_{\theta'\theta}^{(x)}:=\Pr(\theta',x|\theta)$ that
generates the process. By the previous paragraph's conclusion, a merged machine
must have the property that merging its retrodictively equivalent states
results in a co-unifilar machine. 

However, there is more to consider here. We assumed starting with a process' \eM.
This means each state $s_t$ must have a unique prediction of the future
$\Pr\left(x_{t+1}\dots x_{\ell}|s_{t}\right)$. In fact, we can relate these
predictions to the partition:
\begin{align*}
\Pr \left(x_{t+1}\dots x_{\ell}|s_{t}\right)
  & = \mathrm{Tr} \left[K^{(x_{\ell}\dots x_{t+1})}\ketbra{\psi_s}
  K^{(x_{\ell}\dots x_{t+1})\dagger}\right] \\
  & = \sum_{\theta_{t+1}\dots \theta_\ell}
  \Pr(\theta_{t+1},x_{t+1}|\theta(s_t)) \dots
  \Pr(\theta_{\ell},x_{\ell}|\theta_{\ell-1})
  ~.
\end{align*}
Which, we see, only depends on the partition index $\theta$. Then any two
$s_t,s_t'\in\mathcal{B}_\theta$ with $s_t\neq s_t'$ have the same future
prediction, which is not possible for an \eM. Then, it must be the case that
each partition has only one element: $|\mathcal{B}_\theta|=1$ for all $\theta$.
The remainder of the theorem follows directly. $\square$

\subsection{Efficiency of Reverse $q$-Machines}

To establish a similar theorem for the reverse $q$-machine, we must build
similar prerequisite notions. It will be necessary, first, to be very precise
about all the time-reversals at play here.

The reverse $q$-machine is built from the reverse \eM $\mathfrak{G}$ of a
process $\mathrm{Pr}_{\mathfrak{G}}$. This generates strings of random
variables $X_1 \dots X_L$. The time-reversed generator
$\widetilde{\mathfrak{G}}$ is the forward \eM of the process
$\mathrm{Pr}_{\widetilde{\mathfrak{G}}}$ and generates strings of random
variables $\widetilde{X}_1\dots \widetilde{X}_t$, where $\widetilde{X}_j =
X_{t-j+1}$. It is from $\mathfrak{G}$ that we build the reverse $q$-machine.

However, further consider that the process $\mathrm{Pr}_{{\mathfrak{G}}}$ with
random variables ${X}_1\dots {X}_t$ can itself be generated by some forward
\eM, say $\mathfrak{F}=\left(\mathcal{P},\mathcal{X},\{R^{(x)}_{p'p}\}\right)$.
There is a useful result \cite{Elli11a} that we can always write:
\begin{align}
\Pr{}_{\widetilde{\mathfrak{G}}}\left(s_t |{x}_1 \dots {x}_t \right)
  = \sum_{q_t} \Pr_{\mathcal{C}}\left(s_t|q_t\right)
  \Pr{}_{{\mathfrak{F}}}\left(q_t |{x}_1 \dots {x}_t \right)
\end{align}
for some channel $\Pr_{\mathcal{C}}(s_t | q_t)$. Combined with the
synchronization result for \eMs, the following result for reverse \eMs arises
straightforwardly.

\begin{Prop}[Mixed-state synchronization of retrodictors]
Let ${\mathfrak{G}}$ be a reverse \eM and let $\mathfrak{F}$ be the forward \eM
of the same process. For a length-L word ${x}_1 \dots {x}_t$, let
$\hat{p}({x}_1 \dots {x}_t):= \mathrm{argmax}_{p_t}
\Pr{}_{\mathfrak{F}}(q_=t|{x}_1 \dots {x}_t)$. Let $D({x}_1\dots {x}_t) =
\frac{1}{2}\|\Pr_{\mathfrak{G}}(\cdot|{x}_1 \dots
{x}_t)-\Pr_{\mathcal{C}}(\cdot|\hat{p})\|_1$. Then there exist $0<\alpha<1$ and
$K>0$ such that, for all $t$:
\begin{align*}
\Pr\left(D({x}_1\dots {x}_t) >\alpha^t\right) \leq K\alpha^t
  ~.
\end{align*}
This shows that for a sufficiently long word ${x}_1 \dots {x}_t$, the
distribution over states of the reverse \eM converges to one of a finite number
of options (namely, the $\Pr_{\mathcal{C}}(s|q)$ for different $q$).
\end{Prop}
Note that:
\begin{align*}
\Pr{}_{{\mathfrak{G}}}\left(s_t |{x}_1 \dots {x}_t \right) 
    =  \Pr{}_{\mathcal{C}}\left(s_t|\hat{p}\right)
	\Pr{}_{{\mathfrak{F}}}\left(\hat{p} |{x}_1 \dots {x}_t \right)
    + \sum_{p_t\neq \hat{p}}
	\Pr_{\mathcal{C}}\left(s_t|p_t\right)
	\Pr{}_{{\mathfrak{F}}}\left(p_t |{x}_1 \dots {x}_t \right)
  ~.
\end{align*}
So:
\begin{align*}
\Pr{}_{\mathcal{C}}\left(s_t|\hat{p}\right)
  - \Pr\left(s_t|\hat{p}\right)Q({x}_1\dots {x}_t)
  & \leq \Pr{}_{{\mathfrak{G}}}\left(s_t |{x}_1 \dots {x}_t \right) \\
  & \leq \Pr{}_{\mathcal{C}}\left(s_t|\hat{p}\right)
  + \left(1-\Pr\left(s_t|\hat{p}\right)\right)Q({x}_1\dots {x}_t) 
  ~,
\end{align*}
where $Q({x}_1\dots {x}_t) = 1-\Pr{}_{\mathfrak{F}}(p_t|{x}_1 \dots {x}_t)$.
Then the result follows from the forward \eM synchronization theorem. $\square$

Now, we prove a synchronization theorem for the reverse $q$-machine.

\begin{Prop}[Pure-state synchronization of retrodictors]
Let ${\mathfrak{G}}$ be a reverse \eM and let $\mathfrak{F}$ be the forward \eM
of the same process. For a length-L word ${w} = {x}_1 \dots {x}_t$, let
$\hat{p}({x}_1 \dots {x}_t):= \mathrm{argmax}_{p_t}
\Pr{}_{\mathfrak{F}}(p_L|{x}_1 \dots {x}_t)$. Define the state:
\begin{align*}
\ket{{\Psi}_{x_1\dots x_t}}
  := \sum_{s_t} \sqrt{\Pr{}_{\mathcal{C}}\left(s_t|\hat{p}\right)}
  e^{i\phi_{s_t {w}}} \ket{{\psi}_{s_t}}
  ~,
\end{align*}
where $\phi_{s_t w} = \sum_{j=1}^t \phi_{s_j {x}_j}$ (recall each $s_j$ is
determined by $s_{j+1}$ and ${x}_{j+1}$). Last, let ${\rho}_{x_1\dots x_t} :=
\Pr({x}_1\dots {x}_t) {K}^{(x_1\dots x_t)}\rho_\pi {K}^{(x_1\dots x_t)\dagger}$
and let $F({x}_1 \dots {x}_t) = F\left({\rho}_{x_1\dots
x_t},\ketbra{\Psi_{x_1\dots x_t}}\right)$. Then there exist $0<\alpha<1$ and
$K>0$ such that, for all $t$:
\begin{align*}
  \Pr\left(F({x}_1\dots {x}_t) <1-\alpha^t\right) \leq K\alpha^t
  ~.
\end{align*}
This shows that for a sufficiently long word ${x}_1 \dots {x}_t$, the reverse
$q$-machine state converges on a pure state whose amplitudes are determined by
the mapping $\Pr_{\mathcal{C}}(s|p)$ from $\mathfrak{F}$ states to
${\mathfrak{G}}$ states.
\end{Prop}

First, recall that ${x}_1 \dots {x}_t$ corresponds to the reverse of the
sequence $\widetilde{x}_1 \dots \widetilde{x}_t$, generated by the forward \eM
$\widetilde{\mathfrak{G}}$ of the reverse process. We have the relation
$\Pr{}_{\widetilde{\mathfrak{G}}}\left(\widetilde{s}_t|\widetilde{x}_1 \dots
\widetilde{x}_t\right) = \Pr{}_{{\mathfrak{G}}}\left(s_0|{x}_1 \dots
{x}_t\right)$ where $\widetilde{s}_0=s_t$; representing the initial state of
$\mathfrak{G}$ and the final state of $\widetilde{\mathfrak{G}}$ under time
reversal. By the forward \eM synchronization theorem, then the distribution
$\Pr{}_{{\mathfrak{G}}}\left(s_0|{x}_1 \dots {x}_t\right)$ is concentrated on a
particular state $\hat{s}$ with high probability.

Now:
\begin{align*}
{\rho}_{x_1\dots x_t} & = \sum_{\substack{{s}_0\\ {s}_t,{r}_t}}
  \sqrt{\Pr\left({s}_t,{s}_0|{x}_1 \dots {x}_t\right)\Pr\left({r}_t,{s}_0|{x}_1 \dots {x}_t\right)}
  e^{(i\phi_{{s}_t {w}}-i\phi_{{r}_t {w}})}\ket{{\psi}_{{s}_t}}\bra{{\psi}_{{r}_t}} \\
  & = (1-Q)\sum_{{s}_t,{r}_t}
  \sqrt{\Pr\left({s}_t|{x}_1 \dots {x}_t,\hat{s}\right)\Pr\left({r}_t|{x}_1 \dots {x}_t,\hat{s}\right)}
  e^{(i\phi_{{s}_t {w}}-i\phi_{{r}_t {w}})}\ket{{\psi}_{{s}_t}}\bra{{\psi}_{{r}_t}} \\
  & \qquad + Q \sum_{\substack{{s}_0\neq \hat{s}\\ {s}_t,{r}_t}}
  \sqrt{\Pr\left({s}_t|{x}_1 \dots {x}_t,{s}_0\right)\Pr\left({r}_t|{x}_1 \dots {x}_t,{s}_0\right)}
  e^{(i\phi_{{s}_t {w}}-i\phi_{{r}_t {w}})}\ket{{\psi}_{{s}_t}}\bra{{\psi}_{{r}_t}}
  ~,
\end{align*}
where $Q:= 1-\Pr\left(\hat{s}|{x}_1 \dots {x}_t\right)$.

To handle the term $\Pr\left({s}_t|{x}_1 \dots {x}_t,\hat{s}\right)$, we use
the co-unifilarity properties of the reverse \eM. Note that:
\begin{align*}
\Pr\left({s}_t|{x}_1 \dots {x}_t,\hat{s}\right) 
   =  \frac{\Pr\left(\hat{s}|{x}_1 \dots {x}_t,{s}_t\right)\Pr\left({s}_t|{x}_1 \dots {x}_t\right)}{1-Q}
  ~.
\end{align*}
By co-unifilarity, $\Pr\left(\hat{s}|{x}_1 \dots {x}_t,{s}_t\right)$ is
either $0$ or $1$ depending on the value of ${s}_t$. The synchronization
theorem then demands that $\Pr\left({s}_t|{x}_1 \dots {x}_t\right)<Q$ for those
${s}_t$ which assign $\Pr\left(\hat{s}|{x}_1 \dots {x}_t,{s}_t\right)$ a zero
value. Due to this, along with the above equation:
\begin{align*}
  {\rho}_{{x_1\dots x_t}} = \sum_{{s}_t,{r}_t}
  \sqrt{\Pr\left({s}_t|{x}_1 \dots {x}_t\right)\Pr\left({r}_t|{x}_1 \dots {x}_t\right)}
  e^{(i\phi_{{s}_t {w}}-i\phi_{{r}_t {w}})}\ket{{\psi}_{{s}_t}}\bra{{\psi}_{{r}_t}}
  +O(Q)
  ~,
\end{align*}
where $O(Q)$ contains all the terms so far which scale with $Q$. Then
${\rho}_{{x_1\dots x_t}}=\ketbra{\Psi_{{w}}}+O(Q)$. The rest follows as in
Prop. \ref{prop:synch-qm}.
$\square$

With this, we have sufficient information to determine the MCLM of the system
$S_t {A}_t \dots {A}_1$ for sufficiently long $t$. Let $\pi_s$ be the
stationary distribution for $\widetilde{\mathfrak{G}}$'s memory state, and let
$\lambda_q$ be the same for $\mathfrak{F}$. Consider the channel $\Pr(s'|s) =
\sum_q \Pr(s'|q) \Pr(s|q) \lambda_q/\pi_s$. From it we can construct the
ergodic partition $\{\mathcal{B}_\theta\}$ which is defined as the most refined
partition such that $\Pr(s'|s) >0$ only if $\theta(s)=\theta(s')$. This
partition represents information about the state of $\widetilde{\mathfrak{G}}$
which is recoverably encoded in the state of $\mathfrak{F}$. If, on the one
hand, the partition is trivially coarse-grained ($|\{\mathcal{B}_\theta\}|=1$),
then no information about $\widetilde{\mathfrak{G}}$ is recoverably encoded.
If, on the other, the partition is trivially maximal
($|\mathcal{B}_\theta|=1$ for all $\theta$), then the state of
$\widetilde{\mathfrak{G}}$ is actually a function of the state of
$\mathfrak{F}$. Consequently, in this extreme case, $\widetilde{\mathfrak{G}}$
is unifilar in addition to being unifilar.

\begin{Prop}
Let $\rho_{\widetilde{\mathfrak{G}}}(t)$ be the state of the system $A_1\dots
A_t S_t$ at time $t$. Let $\Theta$ be the projective measurement on
$\mathcal{H}_S$ corresponding to the ergodic partition described above. Then,
for sufficiently large $t$, $\Theta$ is the MLCM of $S_t$. Similarly, $X_t
\Theta$ is the MLCM of $A_t S_t$.
\end{Prop}

By Prop. \ref{the:cq_mclm}, the MLCM must leave each $\rho_{x_1\dots x_t}$
unchanged, for all $t$. This is true for $\Theta$. The question is if any
nontrivial refinement, say $Y$, of $\Theta$ can do so. Realize that for
any $\epsilon>0$, there is sufficiently large $t$, so that for each state $s$
there must be at least one word $x_1\dots x_t$ satisfying
$F\left(\rho_{x_1\dots x_t},\ketbra{\Psi_{x_1\dots x_t}}\right)>1-\epsilon$.
Then, for sufficiently large $t$, it must be the case that there exists a word
$x_1\dots x_t$ such that $Y$ modifies $\rho_{x_1\dots x_t}$, because $Y$ (by
virtue of being a refinement of $\Theta$) cannot commute with all the
$\ketbra{\psi_s}$. Therefore, $\Theta$ is the maximal commuting local
measurement. That $X_t \Theta$ is the MLCM of $A_t S_t$ follows from similar
considerations.
$\square$

\begin{The}[Maximally-efficient reverse $q$-machine]
\label{the:quant_retro}
Let $\mathfrak{G}=(\mathcal{S},\mathcal{X},\{T_{s's}^{(x)}\})$ be a given
process' reverse \eM. Suppose we build from it a reverse $q$-machine with
encoding $\left\{\ket{\psi}_s\right\}$ and Kraus operators $\{K^{(x)}\}$. Let
$\mathcal{B}:=\{\mathcal{B}_\theta\}$ be the MCP of $\mathcal{S}$. Then the
quantum generator has $\Delta S_{\mathrm{loc}}=0$ if and only if the partition
$\mathcal{B}$ is trivially maximal---in that $|\mathcal{B}_\theta|=1$ for each
$\theta$---and the predictively state-merged generator
$\mathfrak{G}^{\mathcal{B}}$ of $\mathfrak{G}$ is unifilar.
\end{The}

We define $\rho_{\mathfrak{G}}^{(\theta)}(t): =
\Pi^{(\theta)}\rho_{\mathfrak{G}}(t)\Pi^{(\theta)}$, so that
$\rho_{\mathfrak{G}}(t)=\bigoplus_\theta \rho_{\mathfrak{G}}^{(\theta)}(t)$.
For the equivalence relation we say that $\theta\sim\theta'$ if
$\rho_{\mathfrak{G}}^{(\theta)}(t)$ and $\rho_{\mathfrak{G}}^{(\theta')}(t)$
are unitarily equivalent on the subsystem $S$. This defines a partition
$\mathcal{P}=\left\{[\theta]_{\sim}\right\}$, labeled by $\sigma$, over the set
$\{\mathcal{B}_\theta\}$.

The quantum channel:
\begin{align*}
  \mathcal{E}\left(\rho\right) := \sum_x K^{(x)}\rho K^{(x)\dagger}\otimes \ketbra{x}
\end{align*}
has Kraus operators $L^{(x)}:= K^{(x)}\otimes \ket{x}$. The MLCM of $A_t S_t$ is
$X_t \Theta_t$. Theorem \ref{the:reversible_local} demands that the Kraus
operators $L^{(x)}$ have the form:
\begin{align*}
L^{(x)} & = \bigoplus_{\theta,\theta'}
  \sqrt{\Pr(\theta',x|\theta)} U^{(x, \theta'|\theta)} \\
  & = \bigoplus_{\theta,\theta'} \sqrt{\Pr(\theta',x|\theta)}
  \ket{x}\otimes U_{\theta\mapsto \theta'}^{(x)}
  ~.
\end{align*}
This implies that the quantum generator's Kraus operators have the form:
\begin{align}
K^{(x)} = \bigoplus_{\theta,\theta'}
  \sqrt{\Pr(\theta',x|\theta)} U^{(x)}_{\theta\mapsto \theta'}
  ~.
\label{sm:kraus_form}
\end{align}
The values $\Pr(\theta',x|\theta)$ must be positive only when $\theta' x \sim
\theta$. This would imply that the resulting merged machine is retrodictive.
However, since the states $\mathcal{S}$ are those of the reverse \eM, they
cannot be further merged into a retrodictive machine. It must then be the case
that the partition $\Theta$ is trivially maximal. Consequently, it must be the
case that $\widetilde{\mathfrak{G}}$ is predictive.
$\square$

The consequences of this theorem are augmented by the following statement about
partition $\{\mathcal{B}_\theta\}$ and the stationary distribution $\rho_\pi$.

\begin{Prop}[Block-diagonality of stationary distribution.]
Let $\Theta$ be the projective measurement on $\mathcal{H}_S$ corresponding to
the ergodic partition described above, with projectors $\{\Pi_\theta\}$. Then
$\rho_\pi = \sum_\theta \Pi_\theta \rho_\pi\Pi_\theta$.
\end{Prop}

For, we can express the stationary distribution as:
\begin{align*}
\rho_\pi = \sum_{x_1\dots x_t}\rho_{x_1\dots x_t}
\end{align*}
at arbitrarily long $t$. Taking $t$ sufficiently large, each $\rho_{x_1\dots
x_t}$ is arbitrarily close to $\ketbra{\Psi_{x_1\dots x_t}}$. Since all these
pure states commute with $\Theta$, so does $\rho_\pi$.
$\square$

The primary implication of this theorem, then, is that trivial maximality of
$\Theta$ implies $\rho_\pi$ is diagonal in the basis $\{\ket{\psi_s}\}$, which
itself implies that the reverse $q$-machine does not achieve any nonzero memory
compression.


\begin{thebibliography}{10}

\bibitem{Arut19}
Frank Arute et~al.
\newblock Quantum supremacy using a programmable superconducting processor.
\newblock {\em Nature}, 574(7779):505--510, 2019.

\bibitem{Pedn19}
E.~Pednault, J.~A. Gunnels, G.~Nannicini, L.~Horesh, and R.~Wisnieff.
\newblock {Leveraging Secondary Storage to Simulate Deep 54-qubit Sycamore
  Circuits.}
\newblock 2019.
\newblock arXiv:1910.09534 [quant-ph].

\bibitem{Feyn82a}
M.~Junge, R.~Renner, D.~Sutter, M.~M. Wilde, and A.~Winter.
\newblock Recoverability in quantum information theory.
\newblock {\em Int. J. Theor. Phys.}, 21(6/7), 1982.

\bibitem{Coec16a}
B.~Coecke, T.~Fritz, and R.~W. Spekkens.
\newblock A mathematical theory of resources.
\newblock {\em Info. Comp.}, 250:59, 2016.

\bibitem{Horo09a}
R.~Horodecki, P.~Horodecki, M.~Horodecki, and K.~Horodecki.
\newblock Quantum entanglement.
\newblock {\em Rev. Mod. Phys.}, 81:865, 2009.

\bibitem{Maur93a}
U.~M. Maurer.
\newblock Secret key agreement by public discussion from common information.
\newblock {\em IEEE Trans. Info. Th.}, 39:3, 1993.

\bibitem{Gacs73a}
P.~G{\'a}cs and J.~K{\"o}rner.
\newblock Common information is far less than mutual information.
\newblock {\em Problems of Control and Info. Theory}, 2:119, 1973.

\bibitem{Wyne75a}
A.~D. Wyner.
\newblock The common information of two dependent random variables.
\newblock {\em IEEE Trans. Info. Theory,}, 21(2), 1975.

\bibitem{Kuma14a}
G.~R. Kumar, C.~T. Li, and A.~El Gamal.
\newblock Exact common information.
\newblock 2014.
\newblock arXiv:1402.0062 [cs.IT].

\bibitem{Land61a}
R.~Landauer.
\newblock Irreversibility and heat generation in the computing process.
\newblock {\em IBM J. Res. Develop.}, 5(3):183--191, 1961.

\bibitem{Lloy89a}
S.~Lloyd.
\newblock Use of mutual information to decrease entropy: Implications for the
  second law of thermodynamics.
\newblock {\em Phys. Rev. A}, 39(10):5378, 1989.

\bibitem{Boyd17c}
A.~B. Boyd, D.~Mandal, and J.~P. Crutchfield.
\newblock Correlation-powered information engines and the thermodynamics of
  self-correction.
\newblock {\em Phys. Rev. E}, 95:012152, 2017.

\bibitem{Boyd17b}
A.~B. Boyd, D.~Mandal, and J.~P. Crutchfield.
\newblock Leveraging environmental correlations: The thermodynamics of
  requisite variety.
\newblock {\em J. Stat. Phys.}, 167(6):1555, 2017.

\bibitem{Boyd17a}
A.~B. Boyd, D.~Mandal, and J.~P. Crutchfield.
\newblock Thermodynamics of modularity: Structural costs beyond the {Landauer}
  bound.
\newblock {\em Phys. Rev. X}, 8(3):031036, 2018.

\bibitem{Loom19a}
S.~Loomis and J.~P. Crutchfield.
\newblock Thermal efficiency of quantum memory compression.
\newblock {\em Phys. Rev. Lett.}, to appear, 2019.
\newblock arXiv:1911.00998.

\bibitem{Cove06a}
T.~M. Cover and J.~A. Thomas.
\newblock {\em Elements of Information Theory}.
\newblock Wiley-Interscience, New York, second edition, 2006.

\bibitem{Niel10a}
M.~Nielsen and I.~Chuang.
\newblock {\em Quantum Computation and Quantum Information}.
\newblock Cambridge University Press, New York, 2010.

\bibitem{Petz86a}
D.~Petz.
\newblock Sufficient subalgebras and the relative entropy of states of a von
  neumann algebra.
\newblock {\em Comm. Math. Phys.}, 105(1):123, 1986.

\bibitem{Petz88a}
D.~Petz.
\newblock Sufficiency of channels over von neumann algebras.
\newblock {\em Quart. J. Math. Oxford}, 2(39):97, 1988.

\bibitem{Rusk02a}
M.~B. Ruskai.
\newblock Inequalities for quantum entropy: A review with conditions for
  equality.
\newblock {\em J. Math. Phys.}, 43:4358, 2002.

\bibitem{Hayd04a}
P.~Hayden, R.~Josza, D.~Petz, and A.~Winter.
\newblock Structure of states which satisfy strong subadditivity of quantum
  entropy with equality.
\newblock {\em Comm. Math. Phys.}, 246(2):359, 2004.

\bibitem{Moso04a}
M.~Mosonyi and D.~Petz.
\newblock Structure of sufficient quantum coarse-grainings.
\newblock {\em Lett. Math. Phys.}, 68(1):19, 2004.

\bibitem{Moso05a}
M.~Mosonyi.
\newblock {\em Entropy, Information and Structure of Composite Quantum States}.
\newblock PhD thesis, KU Leuven, 2005.

\bibitem{Petz08a}
D.~Petz.
\newblock {\em Quantum Information Theory and Quantum Statistics}.
\newblock Springer, Berlin, Heidelberg, Springer-Verlag Berlin Heidelberg,
  2008.

\bibitem{Jenc06}
A.~Jen{\v{c}}ov{\'{a}} and D.~Petz.
\newblock Structure of sufficient quantum coarse-grainings.
\newblock {\em Comm. Math. Phys.}, 263:259, 2006.

\bibitem{Lucz13a}
A.~{\L{}}uczak.
\newblock Quantum sufficiency in the operator algebra framework.
\newblock {\em Int. J. Theor. Phys.}, 53:3423, 2013.

\bibitem{Leif14a}
M.~S. Leifer and R.~W. Spekkens.
\newblock A {Bayesian} approach to compatibility, improvement, and pooling of
  quantum states.
\newblock {\em J. Phys. A: Math. Theor.}, 47:275301, 2014.

\bibitem{Baum12a}
B.~Baumgartner and H.~Narnhoffer.
\newblock The structures of state space concerning quantum dynamical
  semigroups.
\newblock {\em Rev. Math. Phys.}, 24(2):1250001, 2012.

\bibitem{Carb16a}
R.~Carbone and Y.~Pautrat.
\newblock Irreducible decompositions and stationary states of quantum channels.
\newblock {\em Reports on Math. Phys.}, 77(3):293, 2016.

\bibitem{Guan18a}
J.~Guan, Y.~Feng, and M.~Ying.
\newblock The structure of decoherence-free subsystems.
\newblock 2018.
\newblock arXiv:1802.04904 [quant-ph].

\bibitem{Albe19a}
V.~V. Albert.
\newblock Asymptotics of quantum channels: conserved quantities, an adiabatic
  limit, and matrix product states.
\newblock {\em Quantum}, 3:151, 2019.

\bibitem{Mand12a}
D.~Mandal and C.~Jarzynski.
\newblock Work and information processing in a solvable model of {Maxwell’s}
  demon.
\newblock {\em PNAS}, 109(29):11641, 2012.

\bibitem{Boyd15a}
A.~B. Boyd, D.~Mandal, and J.~P. Crutchfield.
\newblock Identifying functional thermodynamics in autonomous {Maxwellian}
  ratchets.
\newblock {\em New J. Physics}, 18:023049, 2016.

\bibitem{Garn17b}
A.~J.~P. Garner, J.~Thompson, V.~Vedral, and M.~Gu.
\newblock Thermodynamics of complexity and pattern manipulation.
\newblock {\em Phys. Rev. E}, 95:042140, 2017.

\bibitem{Garn19a}
A.~J.~P. Garner.
\newblock Oracular information and the second law of thermodynamics.
\newblock 2019.
\newblock arXiv:1912.03217 [quant-ph].

\bibitem{Crut88a}
J.~P. Crutchfield and K.~Young.
\newblock Inferring statistical complexity.
\newblock {\em Phys. Rev. Let.}, 63:105--108, 1989.

\bibitem{Shal98a}
C.~R. Shalizi and J.~P. Crutchfield.
\newblock Computational mechanics: Pattern and prediction, structure and
  simplicity.
\newblock {\em J. Stat. Phys.}, 104:817--879, 2001.

\bibitem{Trav12a}
N.~F. Travers and J.~P. Crutchfield.
\newblock Equivalence of history and generator $\epsilon$-machines.
\newblock arxiv.org:1111.4500 [math.PR].

\bibitem{Crut12a}
J.~P. Crutchfield.
\newblock Between order and chaos.
\newblock {\em Nature Physics}, 8:17--24, 2012.

\bibitem{Uppe97a}
D.~R. Upper.
\newblock {\em Theory and Algorithms for Hidden {M}arkov Models and Generalized
  Hidden {M}arkov Models}.
\newblock PhD thesis, University of California, Berkeley, 1997.
\newblock {P}ublished by University Microfilms Intl, Ann Arbor, Michigan.

\bibitem{Gu12a}
M.~Gu, K.~Wiesner, E.~Rieper, and V.~Vedral.
\newblock Quantum mechanics can reduce the complexity of classical models.
\newblock {\em Nature Comm.}, 3(762), 2012.

\bibitem{Maho16a}
J.~R. Mahoney, C.~Aghamohammadi, and J.~P. Crutchfield.
\newblock Occam's quantum strop: Synchronizing and compressing classical
  cryptic processes via a quantum channel.
\newblock {\em Scientific Reports}, 6(20495), 2016.

\bibitem{Riec16a}
P.~M. Riechers, J.~R. Mahoney, C.~Aghamohammadi, and J.~P. Crutchfield.
\newblock A closed-form shave from {Occam's} quantum razor: Exact results for
  quantum compression.
\newblock {\em Phys. Lett. A}, 93:052317, 2016.

\bibitem{Bind17a}
F.~C. Binder, J.~Thompson, and M.~Gu.
\newblock A practical, unitary simulator for {non-Markovian} complex processes.
\newblock {\em Phys. Rev. Lett.}, 120:240502, 2017.

\bibitem{Agha17a}
C.~Aghamohammadi, J.~R. Mahoney, and J.~P. Crutchfield.
\newblock Extreme quantum advantage when simulating classical systems with
  long-range interaction.
\newblock {\em Scientific Reports}, 7(6735), 2017.

\bibitem{Agha18a}
C.~Aghamohammadi, S.~P. Loomis, J.~R. Mahoney, and J.~P. Crutchfield.
\newblock Extreme quantum memory advantage for rare-event sampling.
\newblock {\em Phys. Rev. X}, 8:011025, 2018.

\bibitem{Thom18a}
J.~Thompson, A.~J.~P. Garner, J.~R. Mahoney, J.~P. Crutchfield, V.~Vedral, and
  M.~Gu.
\newblock Causal asymmetry in a quantum world.
\newblock {\em Phys. Rev. X}, 8:031013, 2018.

\bibitem{Loom18a}
S.~Loomis and J.~P. Crutchfield.
\newblock Strong and weak optimizations in classical and quantum models of
  stochastic processes.
\newblock {\em J. Stat. Phys}, 176:1317--1342, 2019.

\bibitem{Liu19a}
Q.~Liu, T.~J. Elliot, F.~C. Binder, C.~Di Franco, and M.~Gu.
\newblock Optimal stochastic modeling with unitary quantum dynamics.
\newblock {\em Phys. Rev. A}, 99:062110, 2019.

\bibitem{Croo08a}
G.~Crooks.
\newblock Quantum operation time reversal.
\newblock {\em Phys. Rev. A}, 77:034101, 2008.

\bibitem{Crut08a}
J.~P. Crutchfield, C.~J. Ellison, and J.~R. Mahoney.
\newblock Time's barbed arrow: {Irreversibility}, crypticity, and stored
  information.
\newblock {\em Phys. Rev. Lett.}, 103(9):094101, 2009.

\bibitem{Elli11a}
C.~J. Ellison, J.~R. Mahoney, R.~G. James, J.~P. Crutchfield, and J.~Reichardt.
\newblock Information symmetries in irreversible processes.
\newblock {\em Chaos}, 21:037107, 2011.

\bibitem{Dahl11a}
O.~C.~O. Dahlsten, R.~Renner, E.~Rieper, and V.~Vedral.
\newblock Inadequacy of {von Neumann} entropy for characterizing extractable
  work.
\newblock {\em New J. Phys.}, 13, 2011.

\bibitem{Ash65a}
R.~B. Ash.
\newblock {\em Information Theory}.
\newblock John Wiley and Sons, New York, 1965.

\bibitem{Pres12a}
J.~Preskill.
\newblock Quantum computing and the entanglement frontier.
\newblock {\em arxiv:1203.5813}.

\bibitem{Trav10a}
N.~Travers and J.~P. Crutchfield.
\newblock Exact synchronization for finite-state sources.
\newblock {\em J. Stat. Phys.}, 145(5):1181--1201, 2011.

\bibitem{Trav10b}
N.~Travers and J.~P. Crutchfield.
\newblock Asymptotic synchronization for finite-state sources.
\newblock {\em J. Stat. Phys.}, 145(5):1202--1223, 2011.

\end{thebibliography}
\end{document}